\documentclass[12pt,preprint]{aastex}
\usepackage{rotating}
\usepackage{graphicx,times}
\usepackage{amsmath}
\usepackage{longtable}
\usepackage{color}

\shorttitle{Ejection Lorentz Factor and Radiation Location of X-ray Flares} 
\shortauthors{Mu et al.}
\begin{document}
\title{The History of GRB Outflows: Ejection Lorentz Factor and Radiation Location of X-ray Flares}
\author{Hui-Jun Mu \altaffilmark{1,2}, Da-Bin Lin\altaffilmark{1,2}, Shao-Qiang Xi \altaffilmark{1, 3}, Ting-Ting Lin\altaffilmark{1,2}, Yuan-Zhu Wang\altaffilmark{1,2},Yun-Feng Liang\altaffilmark{4}, Lian-Zhong L\"{u}\altaffilmark{1,2}, Jin Zhang\altaffilmark{5}, and En-Wei Liang \altaffilmark{1,2,5}}
  \altaffiltext{1}{GXU-NAOC Center for Astrophysics and Space Sciences, Department of Physics, Guangxi University, Nanning 530004, China;lindabin@gxu.edu.cn, lew@gxu.edu.cn}
  \altaffiltext{2}{Guangxi Key Laboratory for the Relativistic Astrophysics, Nanning 530004, China}
  \altaffiltext{3}{Department of Mathematics and Physics, Officers College of CAPF, Chengdu, 610213,China}
  \altaffiltext{4}{Purple Mountain Observatory, Chinese Academy of Sciences, Nanjing 210008, China}
\altaffiltext{5}{National Astronomical Observatories, Chinese Academy of Sciences, Beijing 100012, China}

\begin{abstract}
We present time-resolved spectral analysis of the steep decay segments of 29 bright X-ray flares of gamma-ray bursts (GRBs) observed with the {\em Swift}/X-ray telescope, and model their lightcurves and spectral index evolution behaviors with the curvature effect model.
Our results show that the observed rapid flux decay and strong spectral index evolution with time can be well fit with this model, and the derived characteristic timescales ($t_c$) are in the range of $33\sim 264$ seconds. Using an empirical relation
between the peak luminosity and the Lorentz factor derived from the prompt gamma-rays, we estimate the Lorentz factors of the flares ($\Gamma_{\rm X}$). We obtain $\Gamma_{\rm X}=17\sim 87$ with a median value of $52$, which is smaller than the initial Lorentz factors of prompt gamma-ray fireballs. With the derived $t_c$ and $\Gamma_{\rm X}$, we constrain the radiating regions of 13 X-ray flares, yielding $R_{\rm X}=(0.2\sim 1.1)\times 10^{16}$ cm, which are smaller than the radii of the afterglow fireballs at the peak times of the flares. A long evolution feature from prompt gamma-ray phase to the X-ray epoch is found by incorporating our results with a sample of GRBs whose initial Lorentz factors are available in literatures, i.e., $\Gamma\propto [t_{p}/(1+z)]^{-0.69\pm 0.06}$.  These results may shed lights on the long term evolution of GRB central engines.
\end{abstract}
\section{Introduction} \label{sec:Introduction}
Gamma-ray bursts (GRBs) are the most extreme explosive events in the universe.
The duration of the prompt gamma-rays ranges from milliseconds to thousands of seconds
(\citealp{Kouveliotou1993}),
and afterglows following the prompt gamma-ray phase were found in the X-ray, optical, and radio bands.
It is generally believed that the prompt gamma-rays are from internal shocks of collisions among fireball shells
and the afterglows are from external shocks when the fireball shells propagate into the circum medium
(e.g., \citealp{Meszaros1993}; \citealp{Rees1994}; \citealp{Meszaros1997}; \citealp{Piran2004}; \citealp{Zhang2004}).
With promptly slewing capacity, the X-ray telescope (XRT) onboard the Swift mission observed erratic flares during
the prompt gamma-ray phase and even up to several days post the GRB trigger
(e.g., \citealp{Burrows2005b}; \citealp{Zhang2006}; \citealp{Nousek2006}; \citealp{OBrien2006}; \citealp{Falcone2006};
\citealp{Falcone2007}; \citealp{Chincarini2007}; \citealp{Chincarini2010}).
These flares are found to be internal origin
and they sinal the restart of the GRB central engine after the prompt gamma-rays
(e.g., \citealp{Burrows2005b}; \citealp{Fan2005}; \citealp{Zhang2006}; \citealp{Dai2006};
\citealp{Proga2006}; \citealp{Perna2006}; \citealp{Romano2006}; \citealp{Liang2006}; \citealp{Wu2006} ;
\citealp{Margutti2010}; \citealp{Maxham2009})\footnote{Alternatively, \cite{Hascoet2015} suggested that X-ray flares
may be produced by the long-lived reverse shock when it crosses the tail of the GRB ejecta. Although this scenario may produce the observed temporal feature of the flares, it seems to be difficult in explaining the observed strong spectral variation in the flares (e.g., \citealp{Butler2007}).}.
Taking these flares into account, the duration of the GRB central engines are much longer than the duration
of the prompt gamma-rays (\citealp{Qin2013}; \citealp{Virgili2013}; \citealp{Levan2014}; \citealp{Zhang2014}).

X-ray flares are one of the most powerful diagnostic tools for the GRB central engines,
especially their long term evolution behaviors
(e.g., \citealp{Dai2006}).
The radiation region and the bulk Lorentz factor of fireballs for early prompt gamma-rays
and late X-ray flares are of great theoretical interest.
Although it is still quite uncertain, the radiation region of the prompt gamma-rays
is generally believed to be around $10^{13}-10^{15}$ cm
and the fireballs are ultra-relativistic with a Lorentz factor ($\Gamma_\gamma$)
being greater than 100
(e.g., \citealp{Meszaros1993};  \citealp{Rees1994}; \citealp{Piran2004}; \citealp{Zhang2004}).
With the deceleration timescale observed in the afterglow lightcurves,
the derived $\Gamma_\gamma$ values are usually in the range from 100 to 1000
(\citealp{Sari1999}; \citealp{Kobayashi2007}; \citealp{Molinari2007}),
and they are tightly correlated with the isotropic gamma-ray energy ($E_{\rm iso}$) and luminosity ($L_{\rm iso}$) of the prompt gamma-rays
($\Gamma_\gamma-E_{\gamma, \rm iso}$ relation; Liang et al. 2010, 2013; \citealp{Lv2012}). Furthermore, Liang et al. (2015) discovered a tight $L_{\rm iso}-E_{\rm p, z}-\Gamma_\gamma$ relation, where $E_{\rm p, z}$ is the peak energy of the $\nu f_\nu$ spectrum of prompt gamma-rays in the burst frame.
The value of $\Gamma_\gamma$ may be also estimated with the high energy cutoff in the prompt gamma-ray spectrum
based on ``compactness'' argument (\citealp{Fenimore1993}; \citealp{Woods1995}; \citealp{Baring1997};
\citealp{Lithwick2001}; \citealp{Gupta2008}).
More recently, \cite{Tang2014} systematically searched for such a high energy spectral cutoff/break
in GRBs observed with {\em Fermi}/LAT. They estimated the $\Gamma_\gamma$ values for 9 GRBs with the observed cutoffs by
assuming that these cutoffs are caused by pair-production absorption within the sources.
They found $\Gamma_\gamma$ values are also larger than 100,
and confirmed the $\Gamma_\gamma-E_{\gamma, \rm iso}$ relation.
For late X-ray flares, the Lorentz factor ($\Gamma_{\rm X}$) was suggested to be smaller than that of the fireballs
producing the prompt gamma-rays (e.g., \citealp{Fan2005}; c.f. Burrows et al. 2005).
With the thermal emission observed in X-ray flares,
\cite{Peng2014} obtained $\Gamma_{\rm X}$ being around $60\sim 150$. The curvature effect, which is due to the observer receiving the progressively delayed emission from higher latitudes (\citealp{Fenimore1996};\citealp{Kumar2000}; \citealp{Qin2002}; \citealp{Dermer2004}; \citealp{Uhm2014}), may present tight constraint on the emission region and its Lorentz factor of X-ray flares (e.g., \citealp{Zhang2006}; \citealp{Lazzati2006}). \cite{Jin2010} estimated $\Gamma_{\rm X}$ with both the thermal emission in the flares and the curvature effect on the decay phases of the flares and found that $\Gamma_{\rm X}$ ranges around tens.  The radiation regions of the X-ray flares ($R_{\rm X}$) are even more poorly known. \cite{Troja2014} analyzed the flares with peaking time at $100\sim 300$ seconds post the GRB trigger. They showed that $R_{\rm X}=10^{13}-10^{14}$ cm with $\Gamma_{\rm X}>50$ in the framework of internal shock model if the variability timescale is significantly shorter than the observed flare duration.

As mentioned above, the origin of the flares may be the same as that of prompt gamma-rays.
One may estimate $\Gamma_{\rm X}$ by assuming that the flares follow the same $\Gamma_\gamma-L_{\rm iso}$ relation.
Further more, it is believed that the steep decay observed in the X-ray flares is due to the curvature effect
(Fan \& Wei 2005; \citealp{Dyks2005}; \citealp{Liang2006}; \citealp{Zhang2006}; Panaitescu et al. 2006; \citealp{Wu2006}; \citealp{ZhangBB2007}; \citealp{ZhangBB2009}; \citealp{Qin2008}; c.f., Hasco\"{e}t et al. 2015).
This paper dedicates to study $\Gamma_{\rm X}$ and $R_{\rm X}$ of X-ray flares based the $\Gamma_\gamma-L_{\gamma, \rm iso}$ relation
and the curvature effect on the X-ray flare tails. We select a sample of 29 bright X-ray flares (\S 2)
and fit the lightcurve and the evolving spectral index during the steep decay phases of these X-ray flares based on our curvature effect model (\S 3). We constrain $R_{\rm X}$ and $\Gamma_{\rm X}$ values based on our fitting results in \S 4.
Conclusions and discussions are presented in \S 5. Notation $Q_n=Q/10^{n}$ in cgs units are adopted.

\section{Sample and Data Analysis}\label{Sample Selection}
We present an extensive temporal and spectral analysis for the X-ray flares observed with Swift/XRT during 10 observation years (from 2005 to 2014).
The XRT lightcurve data are downloaded from the website \emph{http://www.swift.ac.uk/} (\citealp{Evans2009}) and fitted
with a multi-component model composing of single power-law and broken power-law functions. This analysis focuses on the steep decay segments of the flares only. We obtain a sample of 29 X-ray flares which satisfy the following criteria. First, they are bright with $F_p/F_u>10$, where $F_p$ and $F_u$ are the peak flux of the flares and the flux of the underlying afterglow component. Second, the decay segments of the flares clearly decline without superimposing other flares or significant fluctuations. The time-resolved spectral analysis for the steep decay segments of flares are made with an absorbed single power-law model,
i.e., $N(E)=N_0*wabs*zwabs*E^{-(\beta+1)}$ and $N(E)=N_0*wabs*wabs*E^{-(\beta+1)}$ with known redshift and unknown redshift,
where ``wabs" and ``zwabs"  are the photoelectric absorbtion of both our Galaxy and GRB host galaxies, respectively.
Since the gas-to-dust ratio of GRB host galaxies are very uncertain
(\citealp{Starling2007}; \citealp{Li2008}; \citealp{Schady2012}),
we ignore the dust scattering effect in calculation of $N_{\rm H}$ values and adopt the same absorbtion model as that of our Galaxy for the GRB host galaxies at redshift $z$.
We derive the $N_{\rm H}$ value of the host galaxy from the X-ray afterglow data for a given burst and
make the time-resolved spectral analysis by keeping this value as a constant.
The spectral analysis results are reported in Table~1. The evolution of the spectral index with time is also shown in Figure~1, where the peak time of the X-ray flare is set as the beginning time ($t=0$) of the steep decay segment for our analysis.

\section{Modeling the Steep Decay Segments in the Curvature Effect Scenario}\label{Sec:Fitting Function}
As mentioned in \S 1, the steep decaying segment of the flares with a slope $\alpha=2+\beta$ would be due to the curvature effect, where $\beta$ is the power-law index of the radiation spectrum (e.g., \citealp{Fenimore1996}; \citealp{Kumar2000}; \citealp{Dermer2004}; \citealp{Liang2006}; \citealp{ZhangBB2007}). The curvature effect is a combination of the time delay and the Doppler shifting of the intrinsic spectrum for high latitude emission with respect to that in the light of sight. The time delay of photons from radius $R_{\rm X}$ and latitude angle $\theta$ with respect to those from $R_{\rm X}$ and $\theta=0$ is given by
\begin{equation}\label{Eq:time}
t = (1 + z)({R_{\rm X}}/c)(1 - \cos\theta ),
\end{equation}
where $z$ is the redshift of the studied source and $c$ is the speed of light. For a relativistic moving jet with a Lorentz factor $\Gamma_{\rm X}$,
the comoving emission frequency $\nu'$ is boosted to $\nu=D\nu'$ in the observer¡¯s frame,
where $D$ is the Doppler factor described as
\begin{eqnarray}
D ={\left[ {{\Gamma _{\rm X}}(1 - {\beta _{{\rm{jet}}}}\cos \theta )} \right]^{ - 1}}
\approx {\left[ {{1 \over {2{\Gamma _{\rm X}}}} + {\Gamma _{\rm X}}(1 - \cos \theta )} \right]^{ - 1}}
=\frac{2\Gamma_{\rm X}}{(1+t/t_c)}
\end{eqnarray}
where $\beta_{\rm jet}c$ is the jet velocity, and $t_c$ is a characteristic timescale of the curvature effect, which is
\begin{equation}\label{Eq:Gamma}
t_c = \frac{R_{\rm X}(1 + z)}{2\Gamma_{\rm X} ^2c}.
\end{equation}
In case of a single power-law radiation spectrum, the observed spectral index would do not evolve with time (e.g., Fenimore et al. 1996; Dermer et al. 2004; Liang et al. 2006). The observed significant spectral softening as shown in Figure 1 would be due to the curvature effect on a curved radiation spectrum (e.g., Zhang et al. 2007; 2009). Following Zhang et al. (2009), we fit the lightcurves and the spectral evolution features of the steep decay segments in our sample in the curvature effect scenario. We take the intrinsic radiation spectrum as a cut-off power-law parameterized as (Zhang et al. 2009)
\footnote{The choice of this function was also due to the fact that the spectral
evolution of some GRB tails can be fitted by such an empirical
model (Campana et al. 2006; Zhang et al. 2007; Yonetoku et al. 2008).},
\begin{equation}
N'(E') = {N'_0}{\left( {\frac{{E'}}{{1{\rm{keV}}}}} \right)^{ -\hat{\beta}}}\exp \left[ { - {{\left( {\frac{{E'}}{{{E'_c}}}} \right)}^\kappa}} \right],
\end{equation}
where $\hat{\beta}$ is the photon index and $E'_c$ is the cut-off energy. Parameter $\kappa$
measures the steepness of the spectrum at $E'>E'_c$,
as shown in Figure 2.
We normally take $\kappa=1$ in this analysis. \cite{ZhangBB2009}
showed that such an intrinsic radiation spectrum can present
the observed spectral index and flux evolution behaviors of GRB 050814 with the curvature effect model.
The observed flux at photon energy $E$ then can be calculated with $F_E\propto D^{2}E'N'(E')$, i.e.,
\begin{equation}
F_E(t) =F_{E,0}\left( {1 + \frac{t}{{{t_c}}}} \right)^{-1- \hat{\beta}}\exp \left[ { - \frac{E}{E_{c,0}}\left( {1 + \frac{t}{{{t_c}}}} \right)} \right]{\left(\frac{E}{1\rm keV} \right) ^{ - \hat{\beta}  + 1}},
\end{equation}
where $F_{E,0}$ and $E_{c,0}\equiv2\Gamma_X E'_c$ are the observed
on-axis flux and cut-off photon energy (corresponding to $t=0$), respectively.
The observed flux in the XRT band then can be given by
\begin{equation}\label{Eq:Flux_Evo}
F_{\rm XRT}(t) = \int_{0.3{\rm{keV}}}^{10{\rm{keV}}} {F_EdE}.
\end{equation}
We simply calculate the observed spectral index in the XRT band at $t$ with
\begin{equation}\label{Eq:Flux_Beta}
\beta (t)  =  - \frac{{\log \left( {{F_{10{\rm{keV}}}}(t)} \right) - \log \left( {{F_{{\rm{0}}{\rm{.3keV}}}}(t)} \right)}}{{\log \left( {10{\rm{keV}}} \right) - \log \left( {{\rm{0}}{\rm{.3keV}}} \right)}}.
\end{equation}

We make jointed fits to the lightcurves and $\beta$ evolution with
Eqs. (\ref{Eq:Flux_Evo}) and (\ref{Eq:Flux_Beta}). The peak time of flares is set as the zero time $t=0$ in these equations
\footnote{Note that our fits by setting $t_0$ as a free parameter may lead to unreasonable results being due to the degeneracy of $t_c$ and $t_0$ in our model. Liang et al. (2006) showed that the $t_0$ values are in the rising segment of the corresponding pulses or flares. Since the flux of flares usually rapidly increase in the rising segment, we simply set the zero time of the flares at the peak times in this analysis.}. The goodness of our fits is evaluated with the total reduced $\chi^2$ by weighting the data points between that in the lightcurves and in the spectral evolution. The total $\chi^2$ value is given by $\chi^2=\chi^2_{\rm LC}+\chi^2_{\beta}$,
where subscripts ``LC" and ``$\beta$" indicate the lightcurves and the $\beta$ evolution curves.
We minimize $\chi^2$ in our fits.

Only 13 GRBs in our sample have redshift measurement.
Our fit curves are displayed in Figure 1, and the model parameters, including $F_{\rm E¡¢ 0}$, $\hat{\beta}$, $E_{c,\ 0}$, and $t_c$, are reported in Table 2.
One can observed that our model can represent the observed flux decay and spectral index evolution in these flares.
The $t_c$ values range from $33$ to $264$ seconds, as shown in Figure 3.

\section{The Lorentz Factor and Radiation Location of X-Ray Flares}
As shown in Eq.~(\ref{Eq:Gamma}), one may estimate $R_{\rm X}$ or $\Gamma_{\rm X}$ with the derived $t_c$ if one of them is available by another way.
By deriving $R_{\rm X}$ with a tentative jet break, \cite{Jin2010} got $\Gamma_{\rm X}=22,13$ for GRBs 050502B and 050724, respectively.
Note that a tight $\Gamma_\gamma-E_{\gamma, \rm iso}$ or $\Gamma_\gamma-L_{\gamma, \rm iso}$ relations were found by \cite{Liang2010} and \cite{Lv2012}\footnote{\cite{Ghirlanda2012} used a different method to estimate the $\Gamma_0$ values. Their method applies the Blandford¨CMckee (BM) self-similar deceleration solution (\citealp{Blandford1976}) and extrapolates it backward to derive $\Gamma_0$. They derived a different slope of the $\Gamma_0-L_{\rm \gamma, \rm iso}$ relation.}. Assuming that the flares follow the $\Gamma_\gamma-L_{\gamma, \rm iso}$ relation, one can use this relation to estimate $\Gamma_{\rm X}$, then derive $R_{\rm X}$ with Eq.~(\ref{Eq:Gamma}).

L\"{u} et al. (2012) derived the $\Gamma_\gamma-L_{\gamma, \rm iso}$ relation by using the average
luminosity of the prompt gamma-rays.
Since burst durations depend on the energy band selected (e. g., Qin et al. 2013),
it is difficult to accurately  measure the duration of flares.
Therefore, we use the peak luminosity of the flares for our analysis.
In addition, L\"{u} et al. (2012) adopted the ordinary least-square regression method to
obtain the regression line.
For regression analysis, the fitting results depend on the
specification of dependent and independent variables (\citealp{Isobe1990}).
One may find discrepancy of the relations among variables by specifying different
dependent variables for a given data set, especially when the data have large error bars or scatters.
To avoid specifying independent and dependent variables in the best linear fits,
we adopt the algorithm of the bisector of two ordinary least-squares to re-do the regression analysis.
The derived $\Gamma_\gamma-L_{\rm \gamma, p}$ relation is
\begin{equation}\label{Eq:Gamma_L}
 \log \Gamma_\gamma=(2.27\pm 0.04)+(0.34\pm 0.03)\log L_{\rm \gamma, p, 52}.
\end{equation}
It is roughly consistent with that reported in L\"{u} et al. (2012).
We check if X-ray flares follow this relation with a flare observed in GRB 050724. With its $L_{\rm X, p}=9.2\times 10^{47}$ erg/s, we get $\Gamma=7.9$, which is comparable to that reported by
\cite{Jin2010}, i.e., $\Gamma_{\rm X}=13$.
We thus use this correlation to estimate the $\Gamma_{\rm X}$ with the peak luminosity of the flares for the 13 GRBs whose redshifts are available.
Our results are reported in Table 3. One can find that $\Gamma_{\rm X}=17\sim 87 $, with a median value of $\Gamma_{\rm X}=52 $.
The derived $\Gamma_{\rm X}$ values are generally consistent with that reported in previous papers
(e.g., \citealp{Fan2005}; \citealp{Falcone2006}; \citealp{Panaitescu2006}).

It is generally believed that X-ray flares are powered by the late activities of the GRB central engine. The derived $\Gamma_{\rm X}$ values in this analysis are systematically smaller than the $\Gamma_0$ values of the prompt gamma-rays as that reported in Liang et al. (2010, 2013) and \cite{Tang2014}. In addition, with the thermal emission observed in the joint BAT and XRT spectra of 13 early flares, which are usually observed at the end of the prompt gamma-ray phase,
\cite{Peng2014} derived the $\Gamma_{\rm X}$ values for these flares
by assuming that the thermal emission is the photosphere emission of the GRB fireballs.
They found that the Lorentz factors range between 50 and 150. They are also systematically larger than the $\Gamma_{\rm X}$ values of late flares in our analysis. To explore possible evolution feature of the Lorentz factor, we plot $\Gamma$ as a function of  $t_p$ in the burst frame for prompt gamma-rays and X-ray flares in Figure~4 with samples from \cite{Liang2013}, \cite{Peng2014}, \cite{Troja2014}, \cite{Tang2014}, and \cite{Fan2005}, where the $t_p$ of prompt gamma-rays is taken as the middle of burst duration.
One can observe a trend that $\Gamma$ values (both $\Gamma_0$ and $\Gamma_{\rm X}$) decay with time. The Spearman correlation analysis yields a correlation coefficient $r=0.70$
and a chance probability $p<10^{-4}$. This may illustrate a long term evolution feature of the GRB central engines (e.g., \citealp{Lazzati2008}). We derive the relation of $\Gamma$ to $t_{\rm p}/(1+z)$ with the algorithm of the bisector of two ordinary least-squares, which yields  $\Gamma=10^{3.05\pm 0.11}\times [t_p/(1+z)]^{-0.69\pm 0.06}$.

With the derived $\Gamma_{\rm X}$, we calculate $R_{\rm X}$ values for the 13 GRBs with Eq. (\ref{Eq:Gamma_L}). The results are also reported in Table 3. It is found that $R_{\rm X}=2.0\times 10^{15}\sim 1.1\times 10^{16}$ cm,
with a median value of $R_{\rm X}=6.5\times10^{15}$ cm. The radiation regions of the flares are within the regions of the prompt
gamma-rays and afterglows. It was generally accepted that the locations of the prompt gamma-rays and afterglows are
$\sim 10^{13}$ cm and $10^{17}$ cm (e.g., \citealp{Zhang2004})
, respectively.
\cite{Troja2014} analyzed the flares with peaking time at $100\sim 300$ seconds post the GRB trigger.
They found $R_{\rm X}\sim 10^{13}-10^{14}$ cm with $\Gamma_{\rm X}>50$ in the framework of internal shock models,
if the variability timescale is significantly shorter than the observed flare duration.
Their $R_{\rm X}$ values are consistent with that producing early prompt gamma-rays.
However, as shown in \cite{Peng2014},
the photosphere radii of the flares at the end of the prompt gamma-ray emission phase
in their sample are usually $10^{13}$ cm.
The internal shock regions of these flares should be beyond the photosphere radius, i.e., $R_{\rm X}\gtrsim 10^{13}$ cm.
The $R_{\rm X}$ values for the late flares in this analysis are in the range of $2.0\times 10^{15}\sim 1.1\times 10^{16}$ cm,
which is much larger than that of the prompt gamma-rays.
We estimate the fireball radii of the afterglows at the flare epochs with $R_{\rm AG}=1.1\times 10^{17} {\rm cm} [(1+z)/2.0]^{-1/2}(E_{\rm k}/10^{52}{\rm erg})^{1/2}A_*^{-1/2}(t_p/1 {\rm day})^{1/2}$ (\citealp{Chevalier2000}), where $E_{\rm k}$ is the kinetic energy of the fireballs and $A_*$ is the wind parameter.
We calculate $E_{\rm k}$ values with $E_{\rm k}=(1-\eta_{\gamma})E_{\gamma,\rm iso}/\eta_{\gamma}$,
where $\eta_{\gamma}$ is the radiative efficiency of prompt emission and it is taken as $0.2$ (\citealp{Frail2001}; \citealp{Molinari2007}; \citealp{Liang2010}). The wind parameter is set as $A_*=1$.
Figure~5 shows the comparison of $R_{\rm X}$ with $R_{\rm AG}$.
We find that they are satisfied with $R_{\rm AG}/50<R_{\rm X}<R_{\rm AG}$, indicating that the emission regions of the flares are in a broad range, but they are smaller than the radii of the fireballs for the afterglows. The narrow distribution of $R_{\rm X}$ of the flares in this analysis would be due to the sample selection effect
since we select only late flares in order to eliminate the contamination of adjacent flares.

\section{Conclusions and Discussion}
We have fit the lightcurves and the spectral evolution during the steep decay segment in 29
late X-ray flares with the curvature effect model.
We show that this model may well represent both the spectral and temporal behaviors in the steep decay segments. The derived characteristic timescale $t_c$ are in the range of $33\sim 264$ seconds, spreading about one order of magnitude. Using the relation between the peak luminosity and the Lorentz factor derived from the prompt gamma-rays, we estimate the Lorentz factors of the flares and find $\Gamma_{\rm X}=17\sim 87$ with a median value of $52$. With the flares in our sample, together with samples collected from literature for prompt gamma-ray emission and early X-ray flares, we find a tentative correlation between the Lorentz factor and the peak time of the flares (or the middle time of the prompt gamma-ray duration), i.e., $\Gamma\propto [t_p/(1+z)]^{-0.69\pm 0.06}$. With the derived $t_c$ and $\Gamma_{\rm X}$, we constrain the radiating region as $R_{\rm X}=2.0\times 10^{15}\sim 1.1\times 10^{16}$ cm, with a median value of $6.5\times 10^{15}$ cm. The radiation regions of the flares are within the regions of the prompt
gamma-rays and afterglows, and the narrow distribution of $R_{\rm X}$ of the flares in this analysis would be due to the sample selection effect
since we select only late flares in order to eliminate the contamination of adjacent flares.

The derived $\Gamma-t_{\rm p}$ anti-correlation indicates the decay of the Lorentz factor of ejecta. It may feature the long-term evolution of central engines.
With the relation of $\Gamma\propto [t_p/(1+z)]^{-0.69\pm 0.06}$ and Eq.~(\ref{Eq:Gamma_L}),
one can obtain $L_{\rm X, p}\propto [t_p/(1+z)]^{-2}$. Lazzati et al. (2008) derived a similar power-law index for the correlation between the mean luminosity and the peak time of the individual flares for 10 long GRBs that have multiple flares.
They suggested that accretion onto a compact object could explain this feature. Statistics for bright flares gives $L\propto [t_{\rm p}/(1+z)]^{-1.9}$ (Chincarini et al. 2010). Margutti et al. (2011) found the average peak luminosity of the early flares ($30 <t_p< 1000$ seconds) decays as $L\propto t^{-2.7\pm 0.1}$.
They argued that this feature could be triggered by a rapid outward expansion of
an accretion shock in the material feeding a convective disc in hyper-accreting black hole scenario.
Note that the discrepancy of the derived power-law indices would be due to the sample selection.

Note that the luminosity $L_{\rm X, p}$ values of the flares in our work are calculated in the XRT band. They may be underestimated, especially for the early X-ray flares. It is well known that the GRB spectra are usually well fit with the Band function in a broad energy band (e.g., Zhang et al. 2011). As shown in Peng et al. (2014), some early X-ray flares are the soft extension of the gamma-ray pulses and the $E_p$ values of their spectra are higher than the XRT band. Their X-ray luminosity values observed in the XRT band are only a small fraction of their bolometric luminosity. Looking at Table 1, one can observe $\beta<1$ for some flares, indicating that their energy fluxes would go up to a higher energy band. For those flares with $\beta>1$, the luminosity observed in the XRT band could be a good representative of the bolometric luminosity, and the $\Gamma_{\rm X}$ values derived from Eq. (8) with $L_{\rm X, p}$ measured in the XRT band is only lower limits of $\Gamma_X$. Then, the true $\Gamma-t_{\rm p}$ relation would become shallower than that derived in this analysis. In addition, with spectral information of prompt gamma-rays collected from Butler et al. (2007), Ghirlanda et al. (2008),and Heussa et al. (2013), we also derive the luminosity at 10 keV ($L_{\rm 10\ keV}$) for the GRBs reported in LV (2012) and checked whether it is still tightly correlated with $\Gamma_0$. We found that it is not. This is reasonable since $L_{\rm 10\ keV}$ is only a very small fraction of the radiation luminosity. Although $\Gamma_0$ is tightly correlated with the bolometric luminosity, it is not necessary to be correlated with any selected mono-frequency luminosity. Therefore, in case of that the luminosity in the XRT band is not a good representative of the bolometric luminosity, the derived $\Gamma_X$ with Eq. (8) may be quite uncertain.

The $R_{\rm X}$ values derived in this analysis are between the prompt gamma-ray and afterglow radiating regions.
 Since $\Gamma_{\rm X}$ may be underestimated as we mentioned above, the inferred $R_{\rm X}$ with Eq. (3) thus ma ybe also underestimated, which is somewhat compensated by the assumption that the flare decay timescale
$t_c$ is set by the curvature effect. It could be that $t_c$ is dominated by something else, and the $t_c$ reported
here may overestimate the true curvature timescale. Recently, Uhm \& Zhang (2015) found that the decay slope is steeper than the standard value from the curvature effect model if the jet is undergoing bulk acceleration when the emission ceases. They showed that the decay properties of flares demand that the emission region is undergoing significant bulk acceleration (seel also Jia et al. 2016). Therefore, the dynamical timescale and the magnetic field decay timescale would dominate the flare decay timescale and the $t_c$ value may be smaller, hence the real $R_{\rm X}$ would be smaller than that reported in this analysis.

\acknowledgments
We acknowledge the use of the public data from the Swift data archive.
We thank the anonymous referee for helpful suggestions to improve the paper.
We also appreciate helpful discussion with Bing Zhang, Xue-Feng Wu, and Shu-Jin Hou.
This work is supported by the National Basic Research Program of China (973 Program, grant No. 2014CB845800),
the National Natural Science Foundation of China (Grant No. 11533003, 11403005, 11373036, 11163001),
the Strategic Priority Research Program ¡°The Emergence of Cosmological Structures¡± of the Chinese Academy of Sciences (grant XDB09000000), the Guangxi Science Foundation (Grant No. 2014GXNSFBA118004, 2013GXNSFFA019001), and the Project Sponsored by the Scientific Research Foundation of Guangxi University (Grant No. XJZ140331).

\begin{figure*}
\includegraphics[angle=0,scale=0.350,width=0.5\textwidth,height=0.25\textheight]{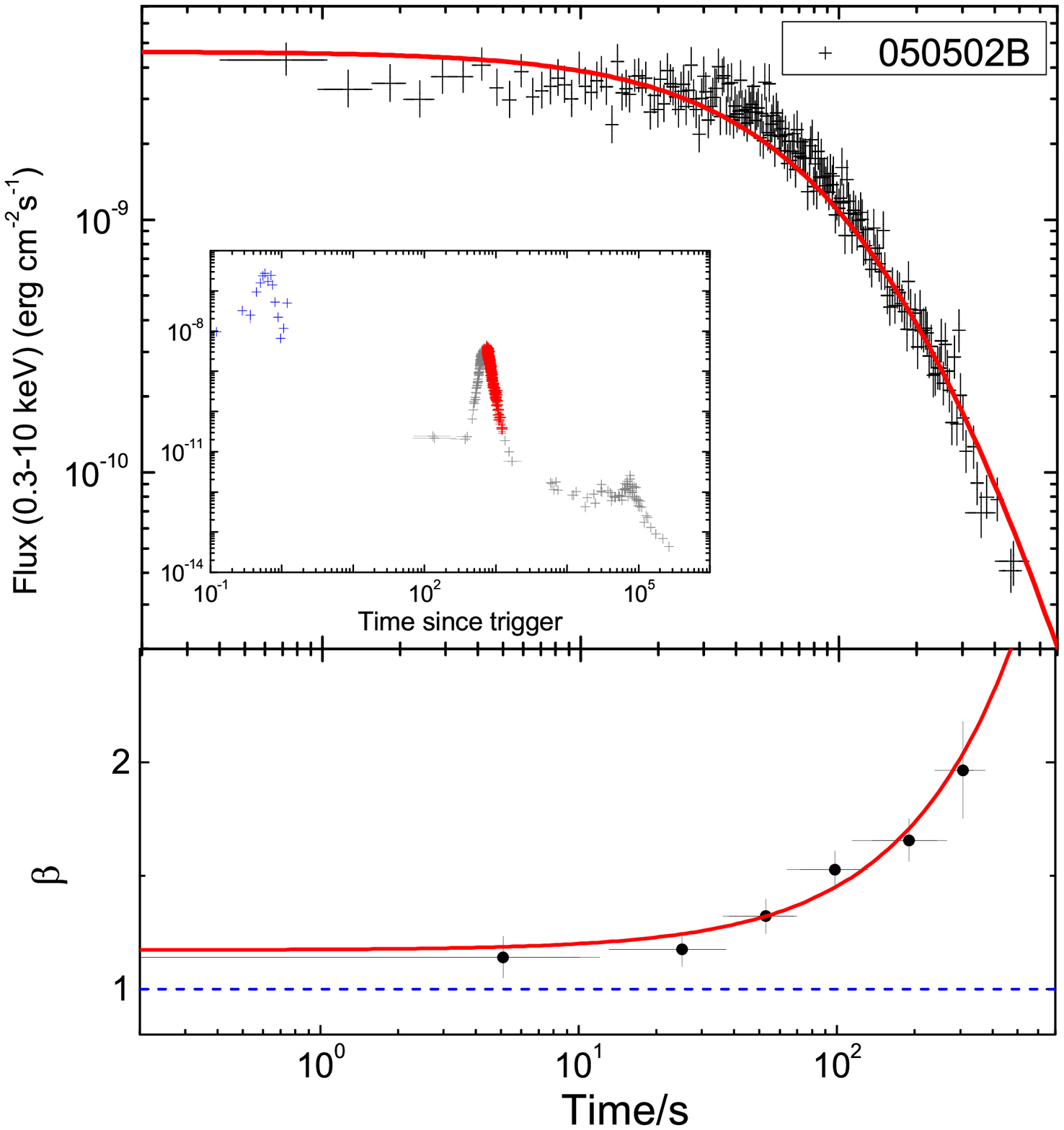}%
\includegraphics[angle=0,scale=0.350,width=0.5\textwidth,height=0.25\textheight]{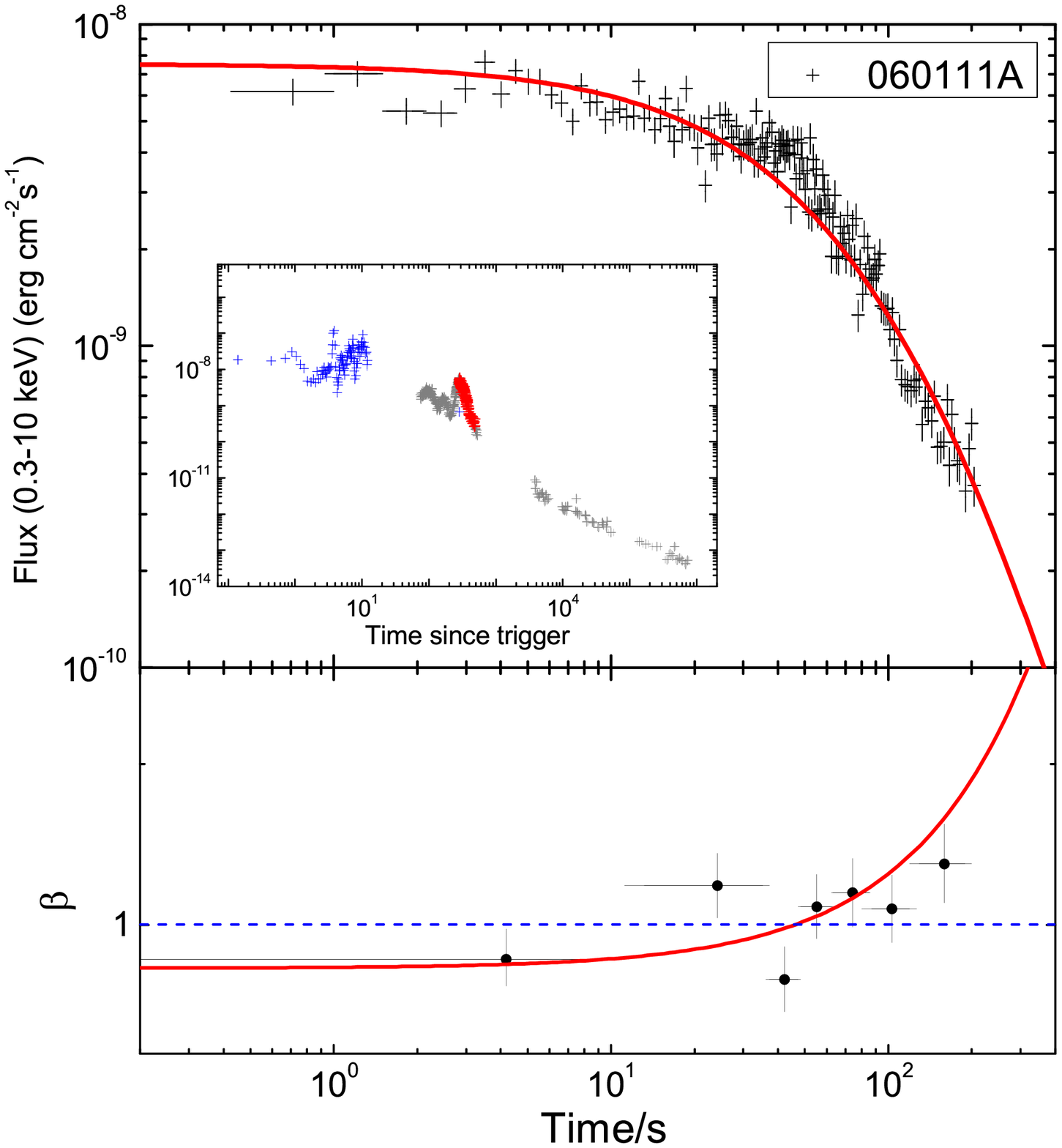}
\includegraphics[angle=0,scale=0.350,width=0.5\textwidth,height=0.25\textheight]{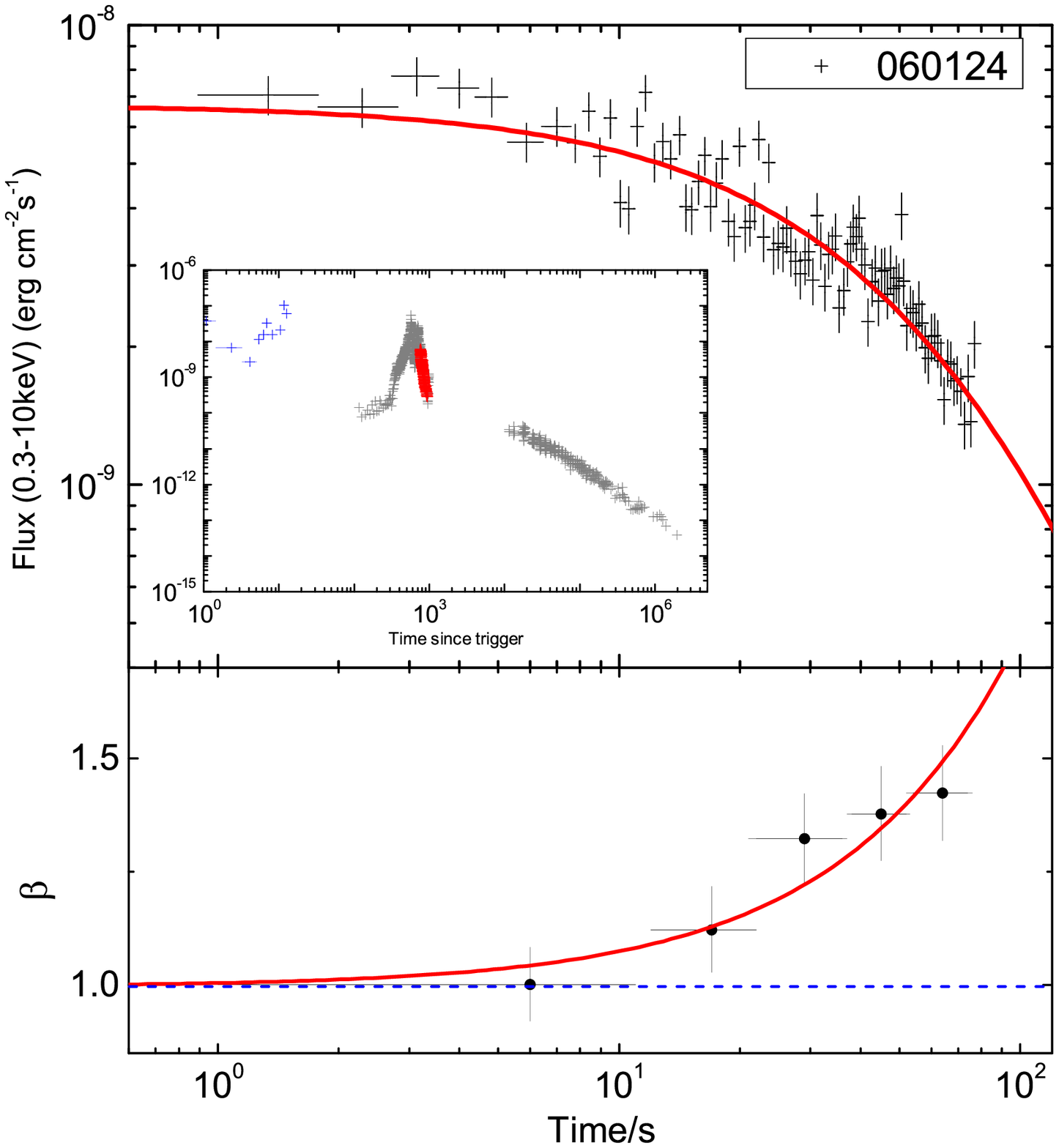}%
\includegraphics[angle=0,scale=0.350,width=0.5\textwidth,height=0.25\textheight]{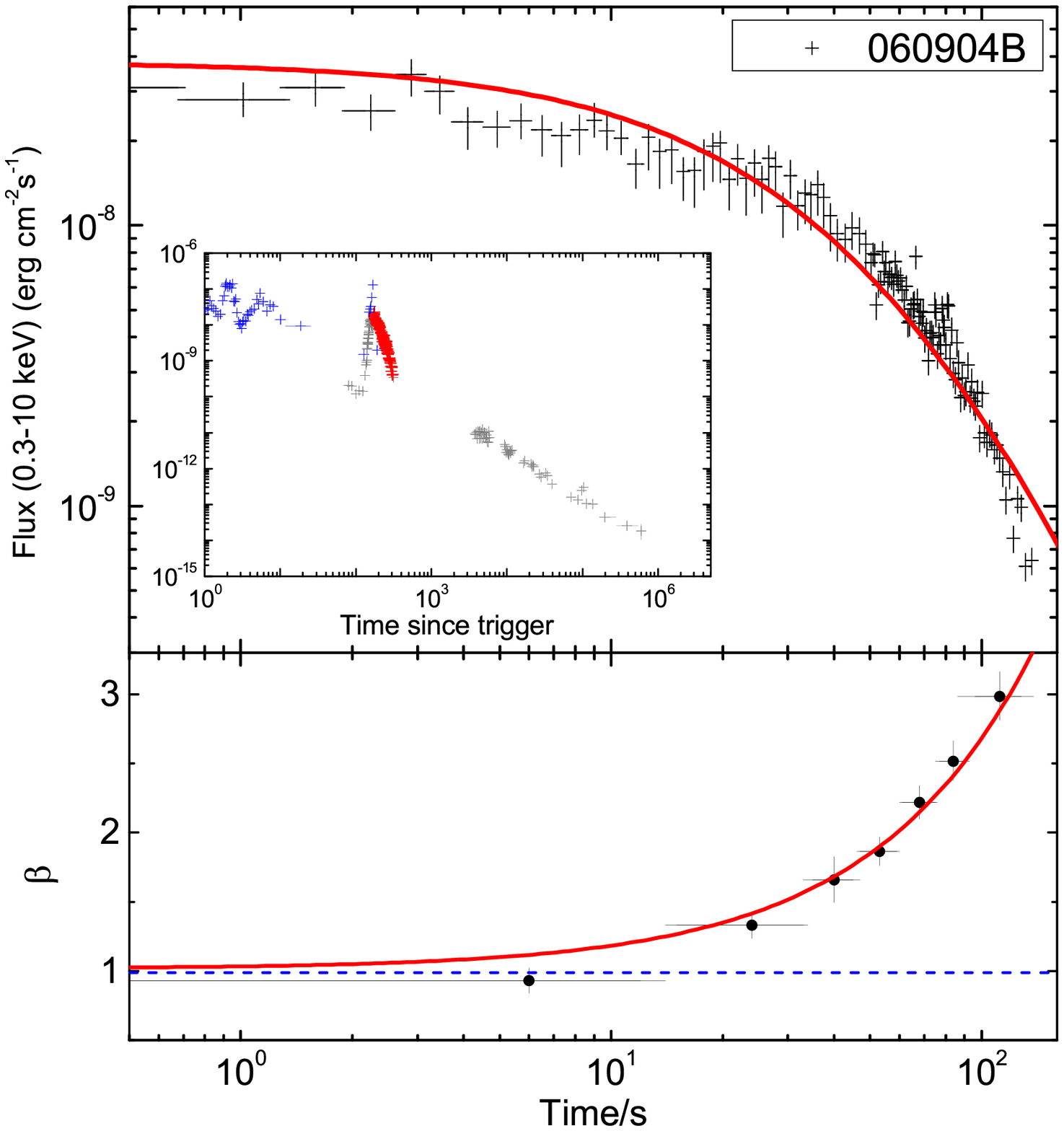}
\includegraphics[angle=0,scale=0.350,width=0.5\textwidth,height=0.25\textheight]{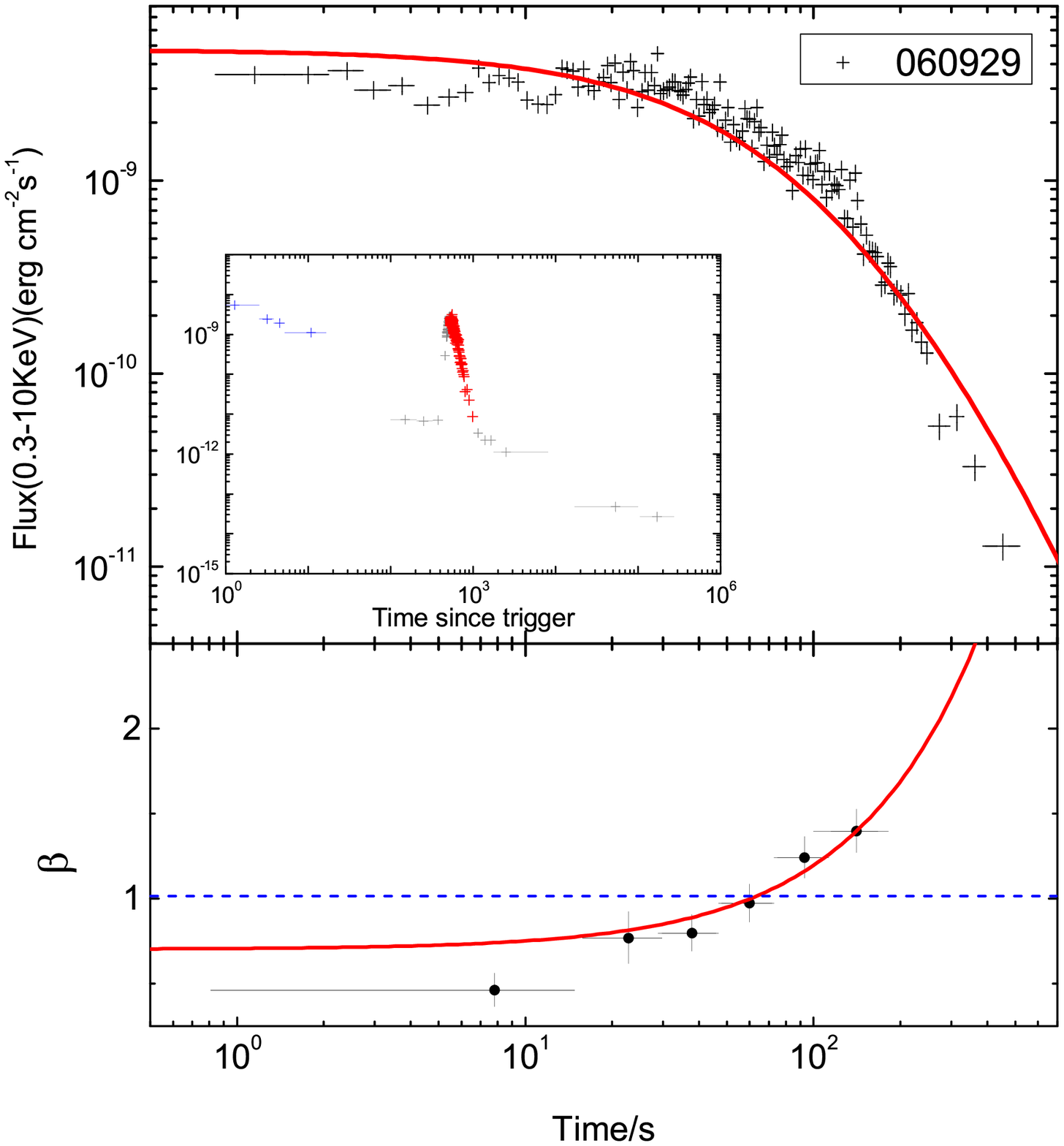}%
\includegraphics[angle=0,scale=0.350,width=0.5\textwidth,height=0.25\textheight]{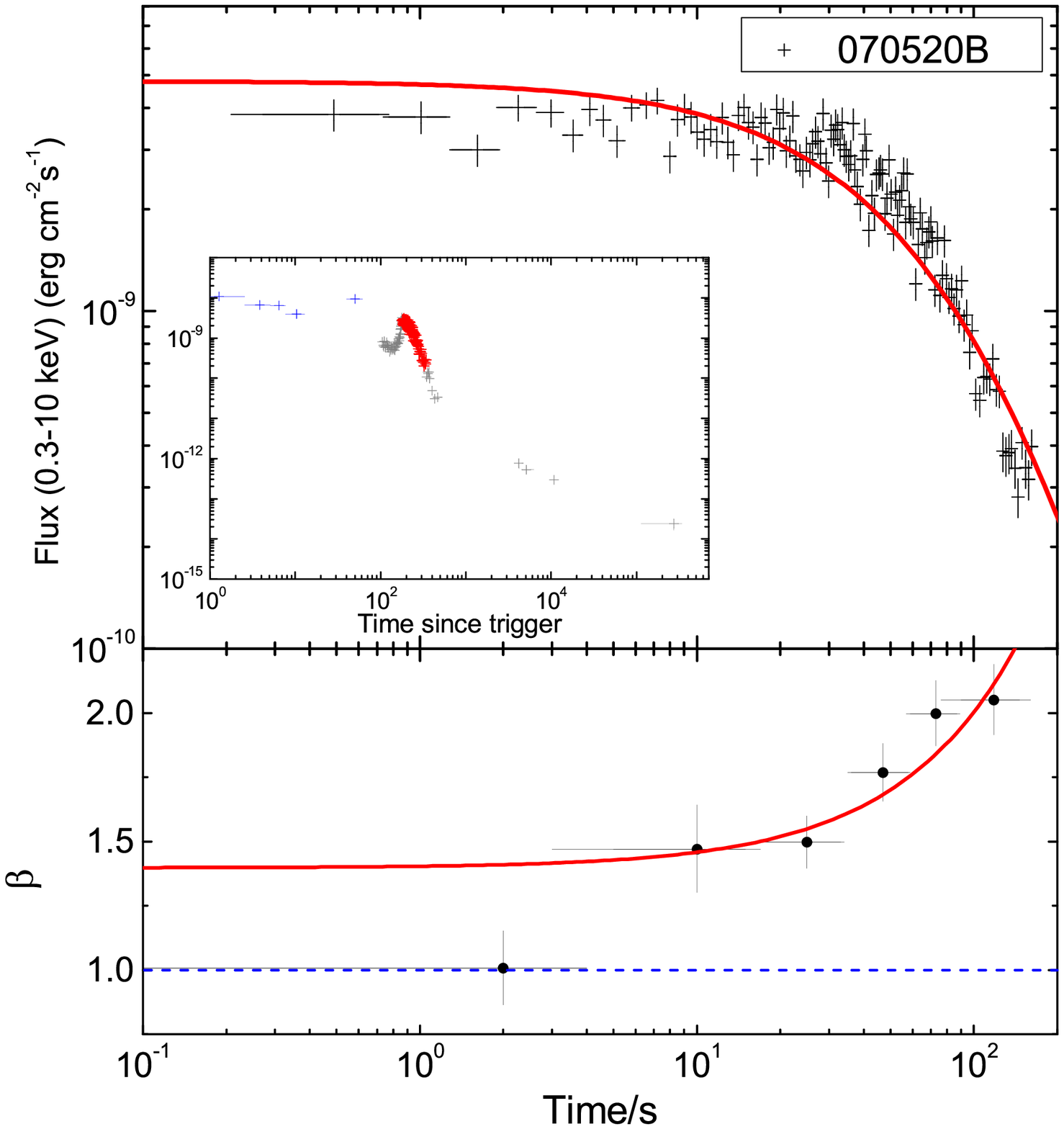}
\includegraphics[angle=0,scale=0.350,width=0.5\textwidth,height=0.25\textheight]{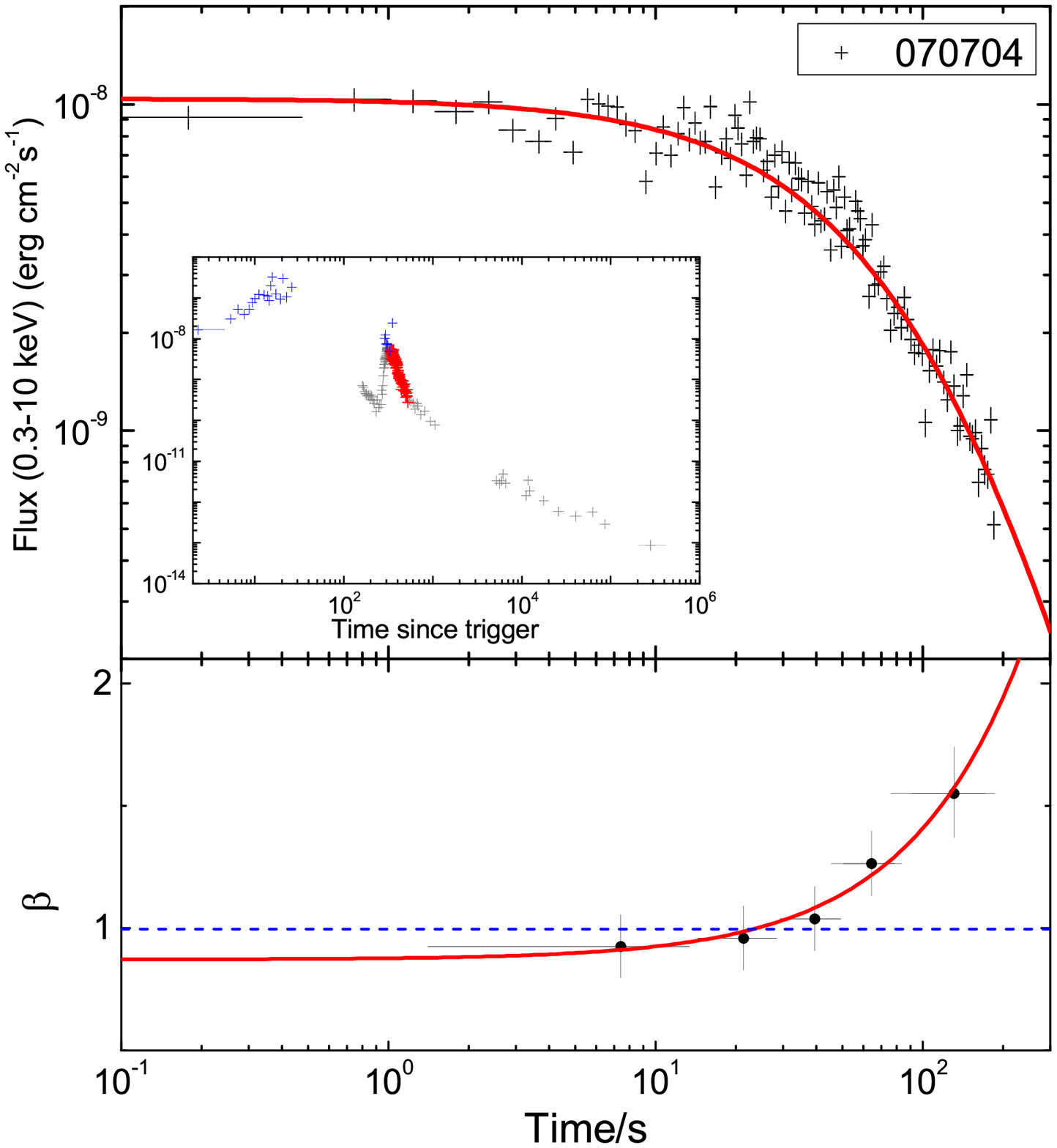}%
\includegraphics[angle=0,scale=0.350,width=0.5\textwidth,height=0.25\textheight]{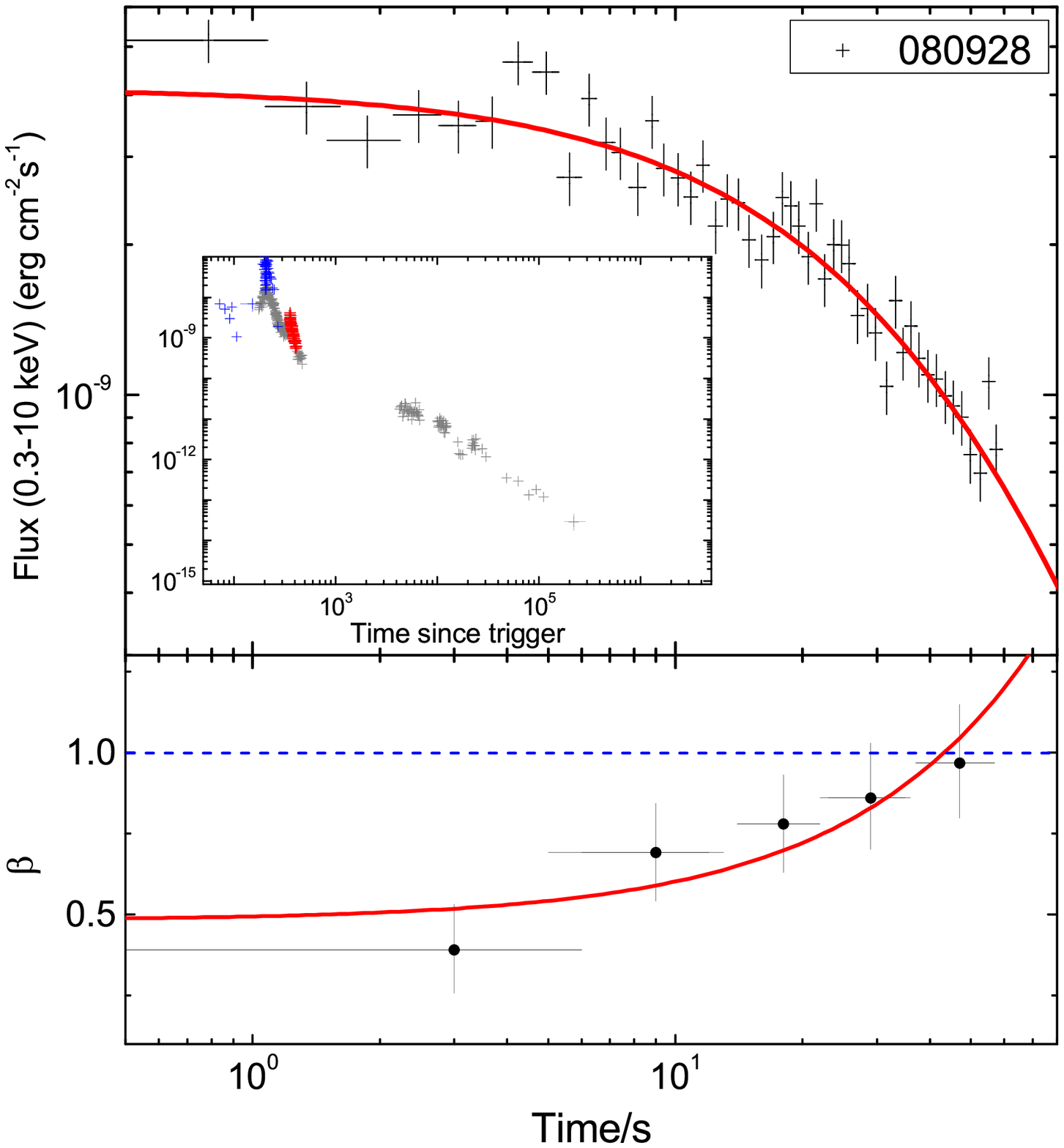}

\caption{Lightcurves (upper part of each panel) and spectral index $\beta$ evolution (lower part of each panel)
of the steep decay segment in our selected 29 flares.
The zero time is set to the peak time of the flares to timing the flare tails.
Our joint fits are also shown with solid curves. The inset shows the light curve in $0.3 - 10$ keV of XRT observation (red and gray) and extrapolation based on BAT observation (blue),
where the red are the data we fitted. The blue dotted line is $\beta=1$.}
\end{figure*}
 \addtocounter{figure}{-1}
\begin{figure*}
\includegraphics[angle=0,scale=0.350,width=0.5\textwidth,height=0.25\textheight]{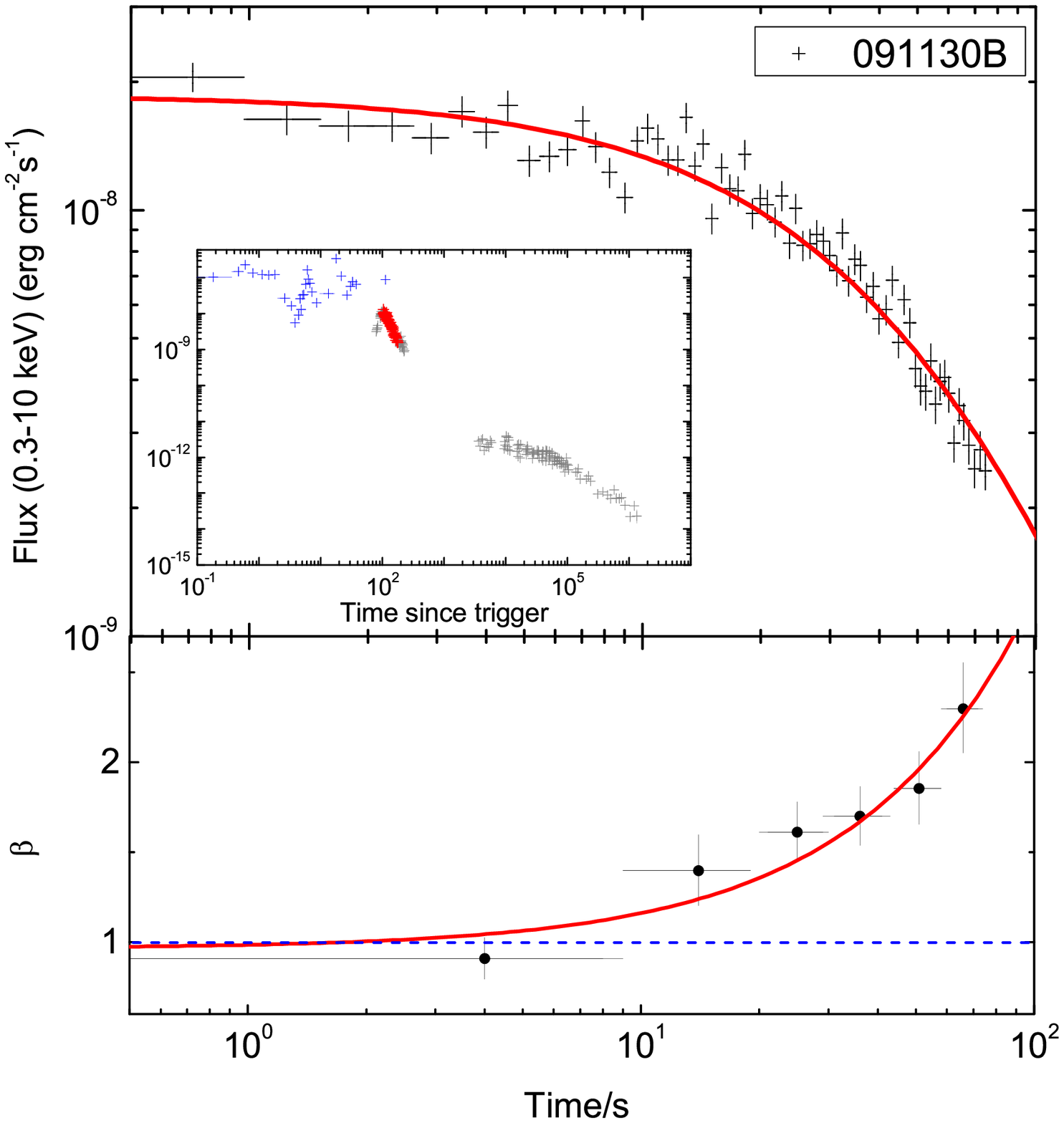}%
\includegraphics[angle=0,scale=0.350,width=0.5\textwidth,height=0.25\textheight]{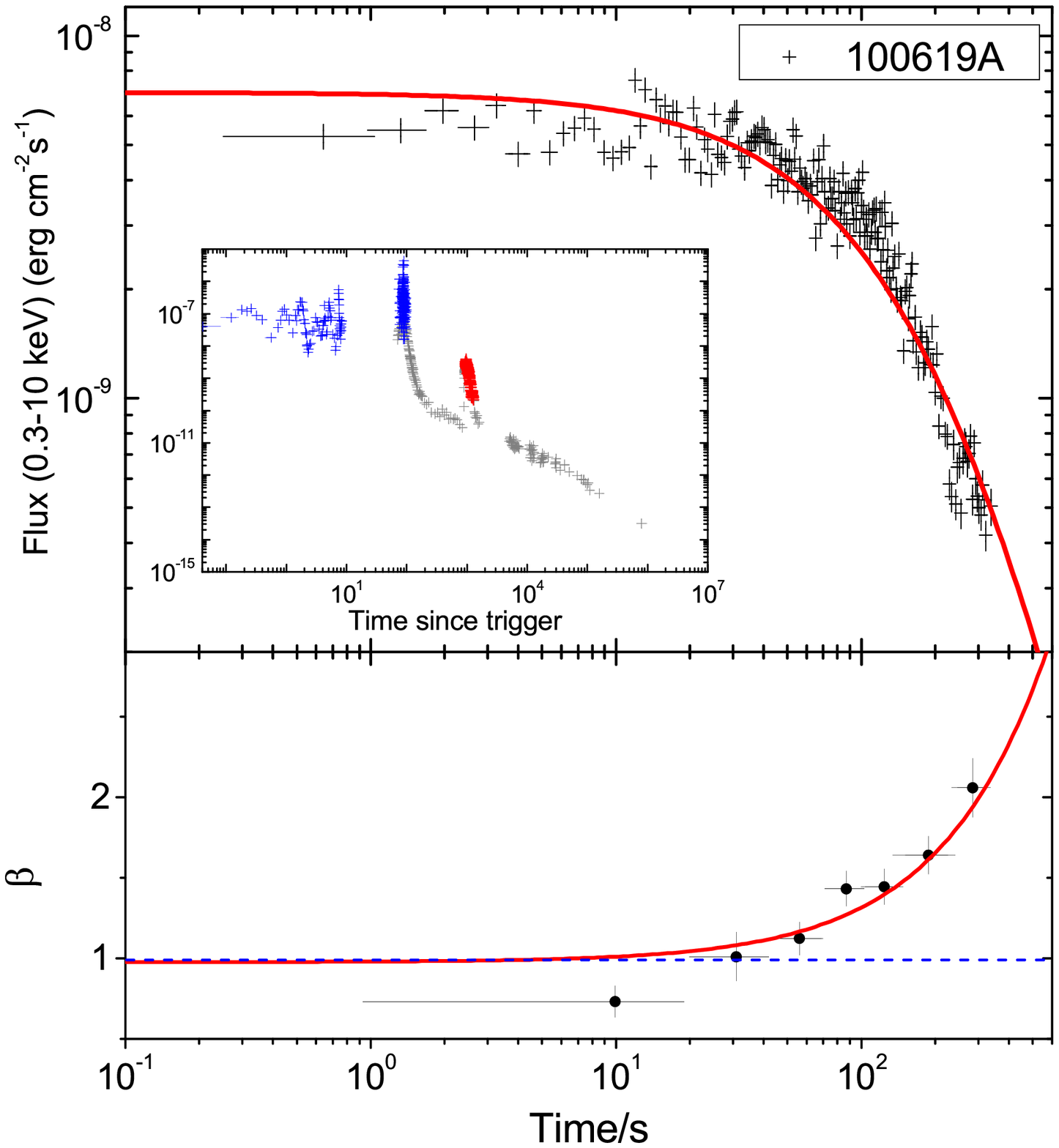}
\includegraphics[angle=0,scale=0.350,width=0.5\textwidth,height=0.25\textheight]{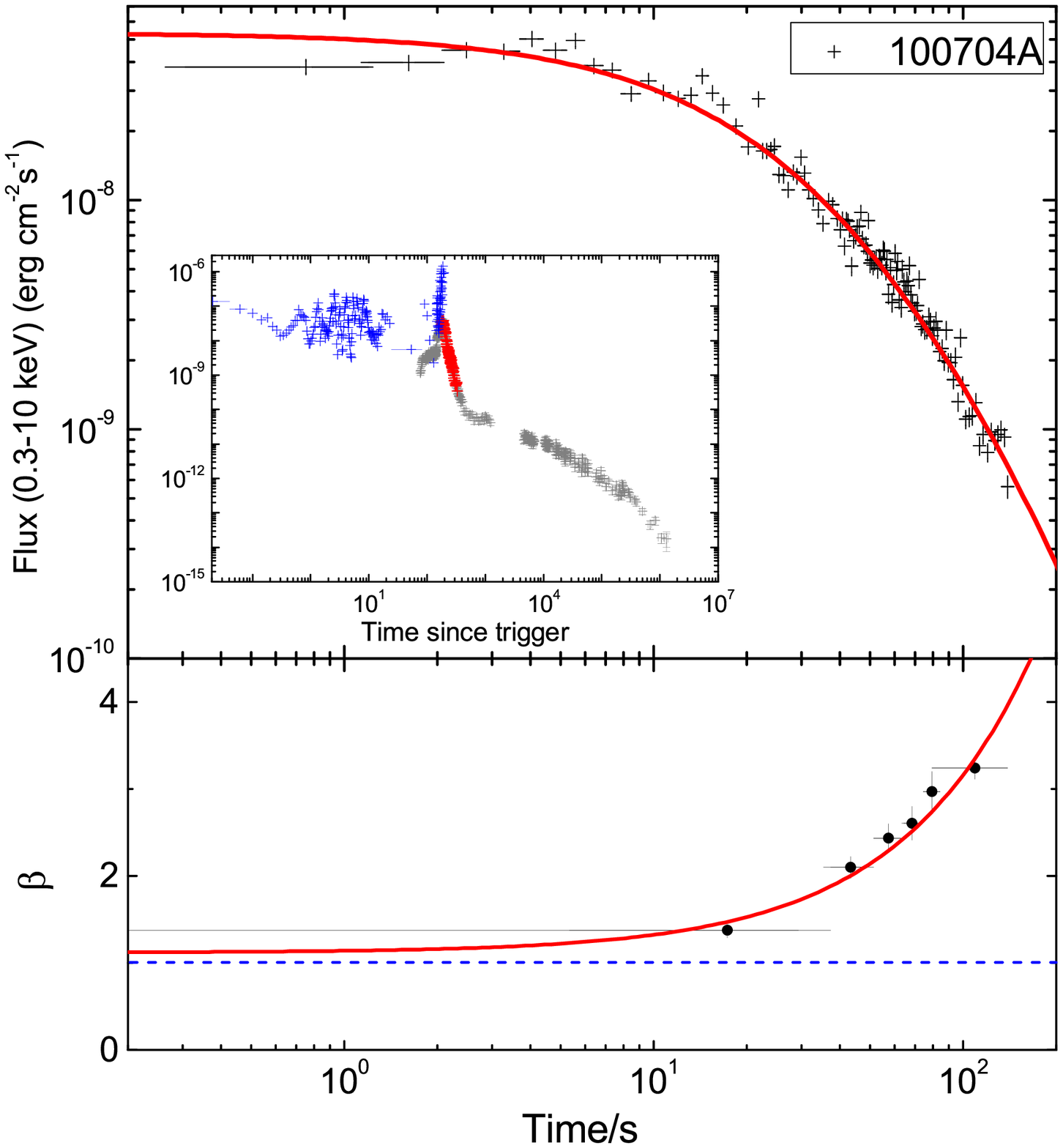}%
\includegraphics[angle=0,scale=0.350,width=0.5\textwidth,height=0.25\textheight]{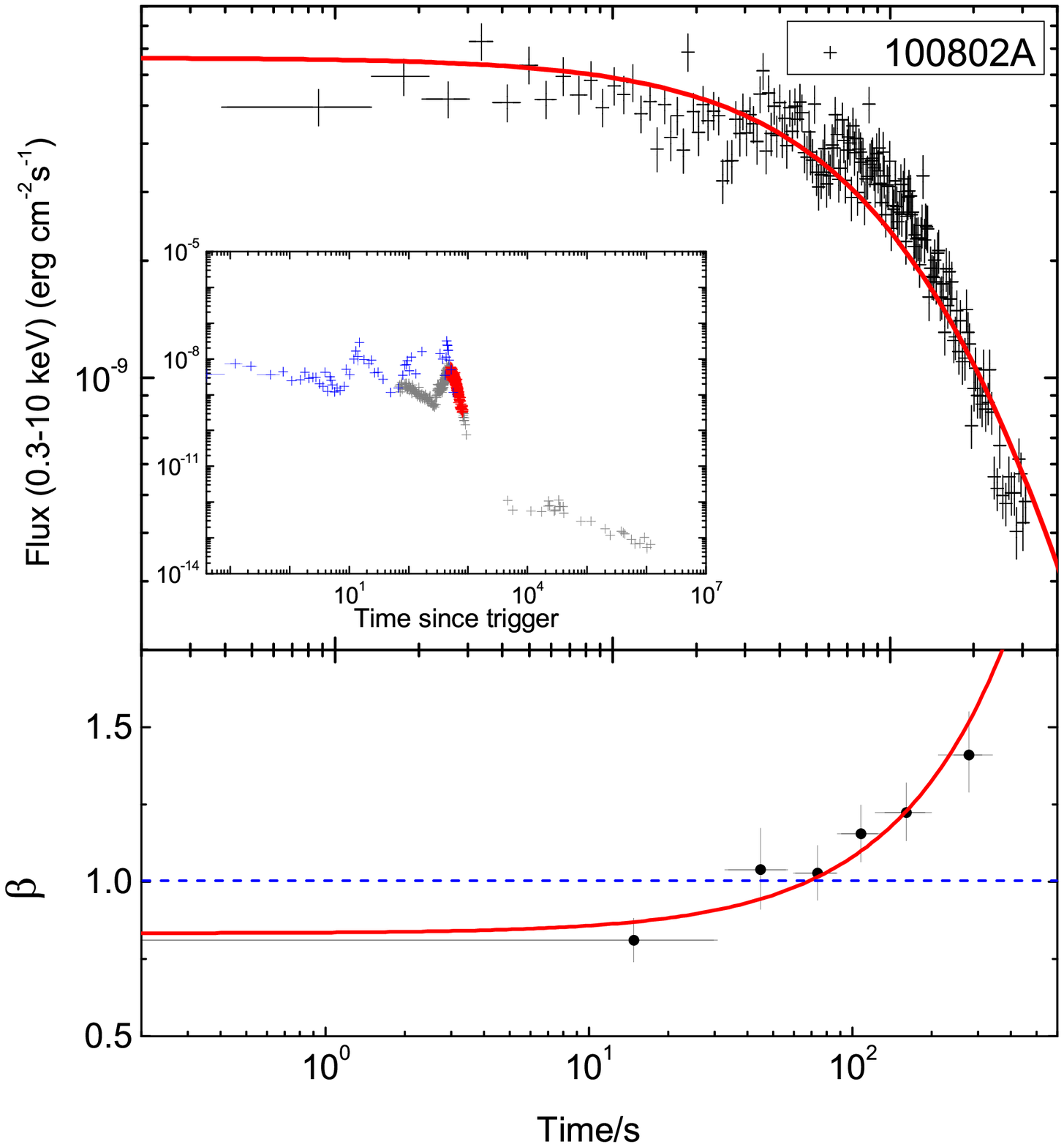}
\includegraphics[angle=0,scale=0.350,width=0.5\textwidth,height=0.25\textheight]{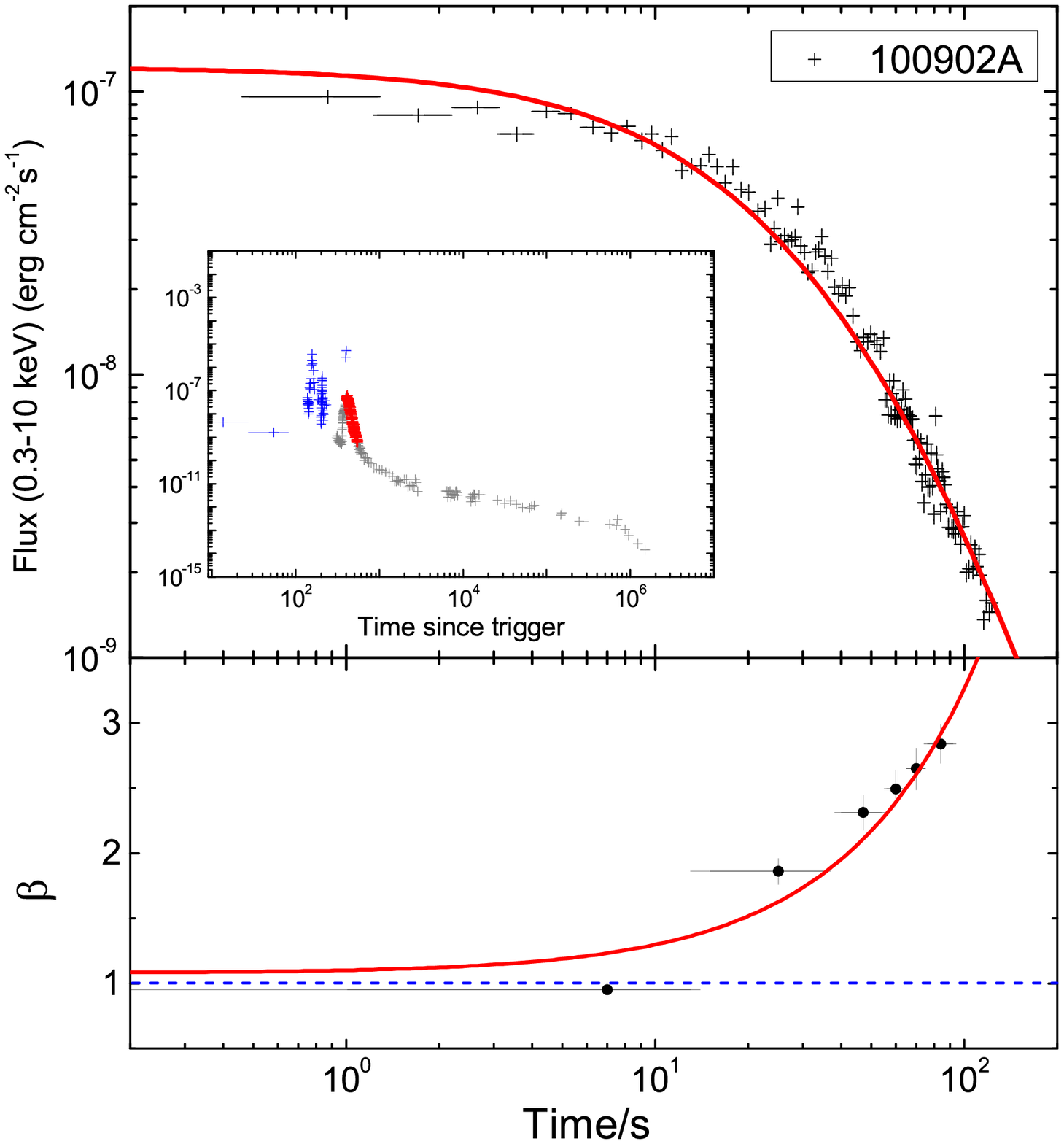}%
\includegraphics[angle=0,scale=0.350,width=0.5\textwidth,height=0.25\textheight]{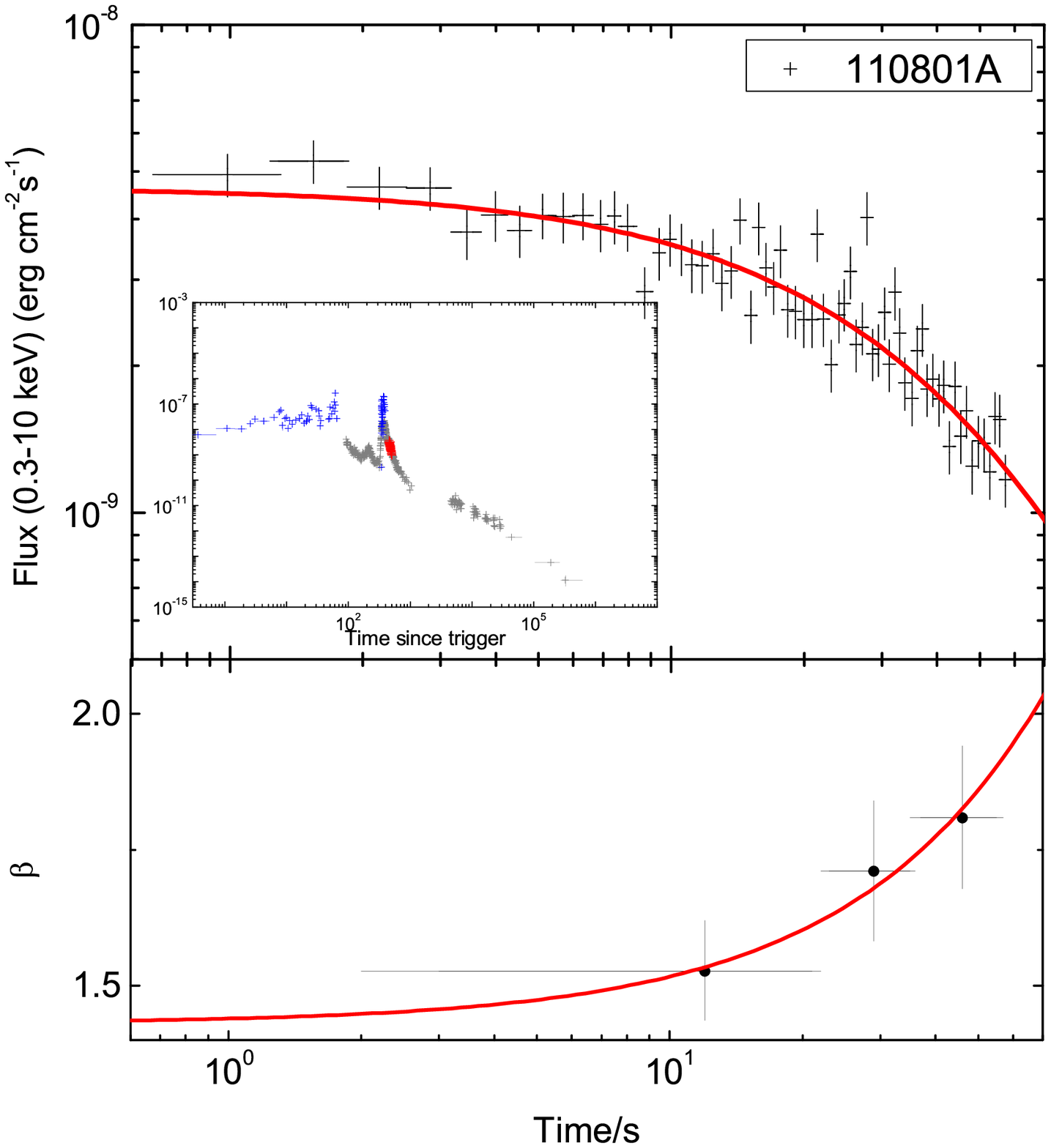}
\includegraphics[angle=0,scale=0.350,width=0.5\textwidth,height=0.25\textheight]{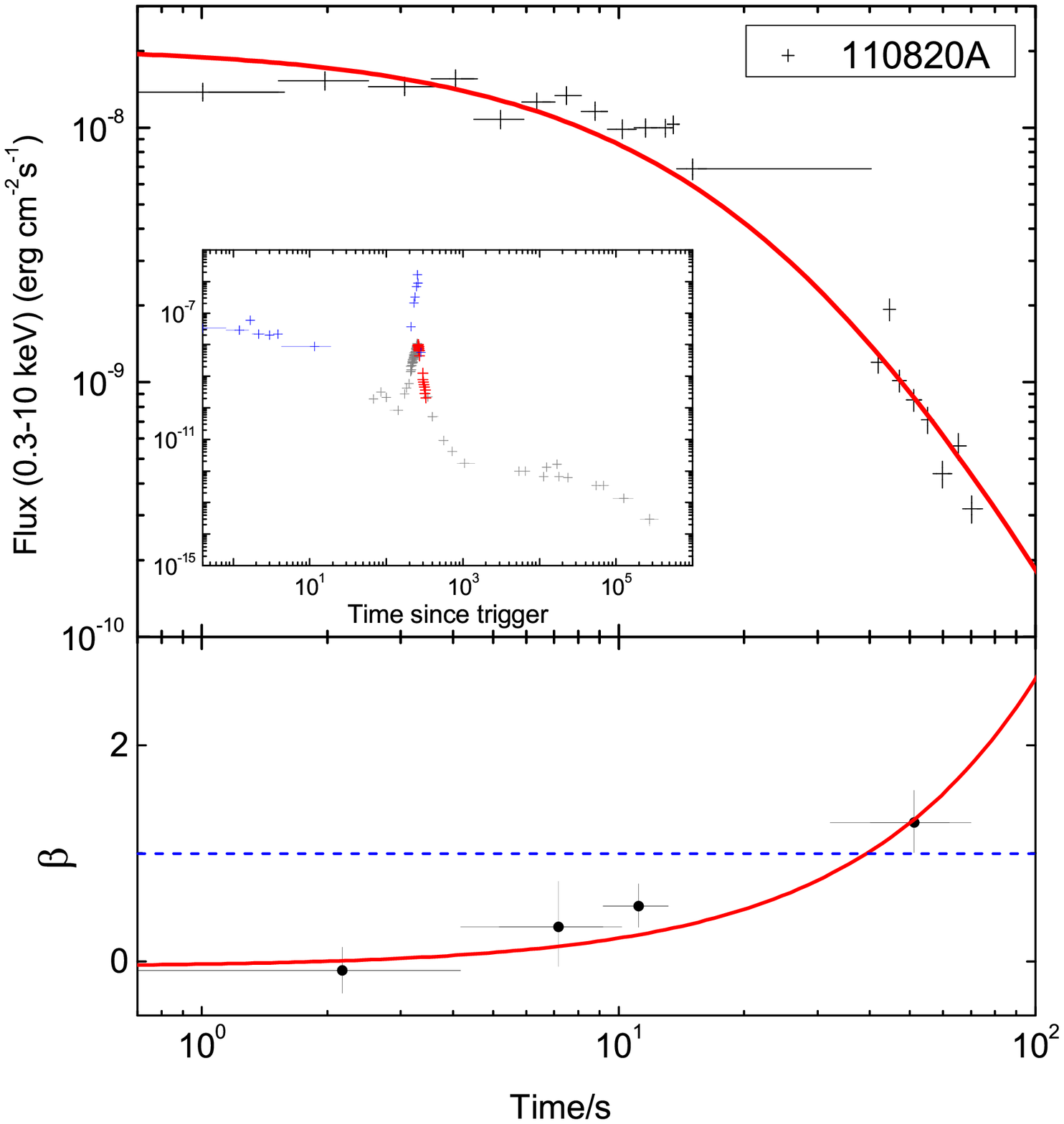}%
\includegraphics[angle=0,scale=0.350,width=0.5\textwidth,height=0.25\textheight]{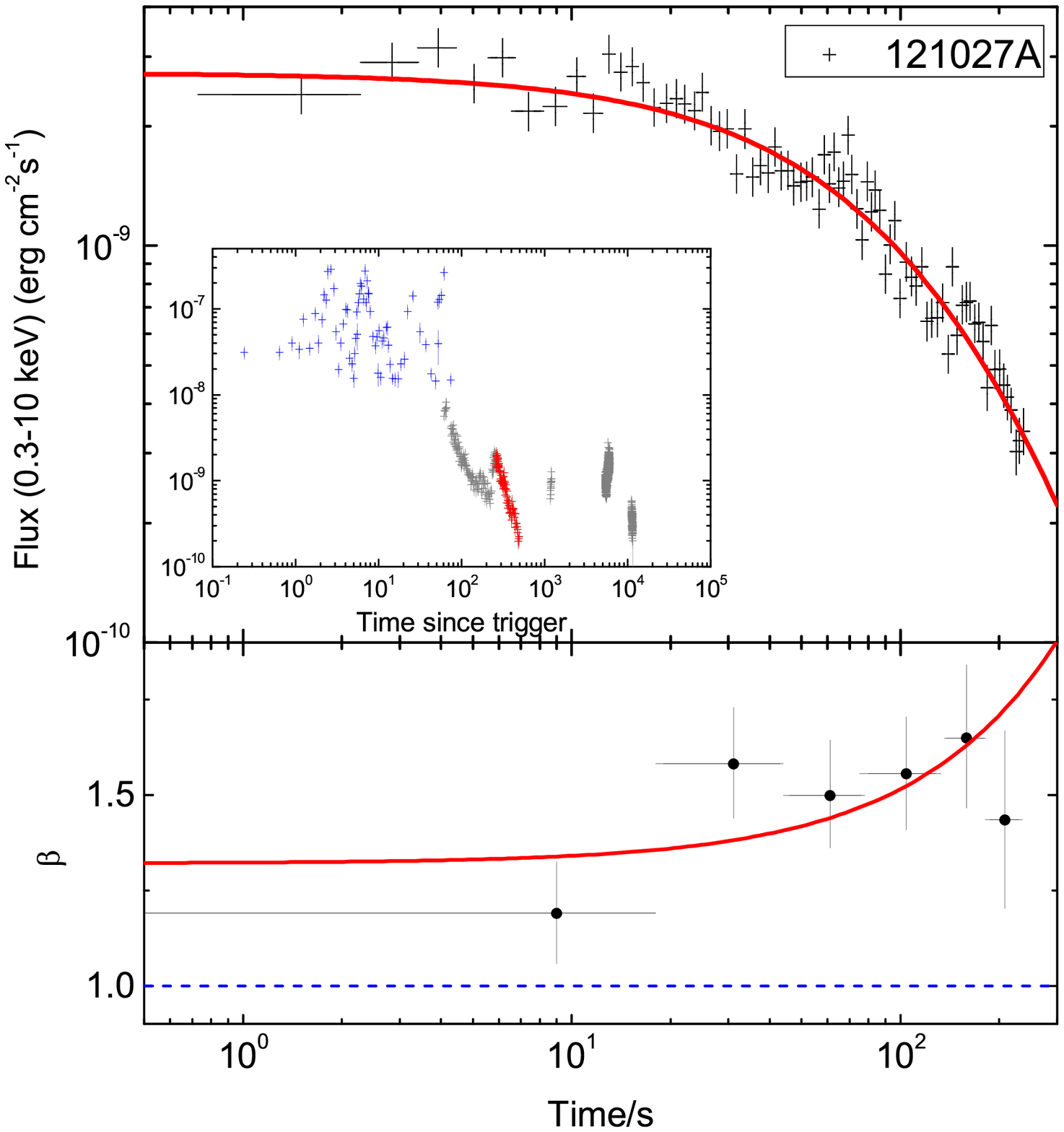}

\caption{(Continued)}
\end{figure*}
 \addtocounter{figure}{-1}
\begin{figure*}
\includegraphics[angle=0,scale=0.350,width=0.5\textwidth,height=0.25\textheight]{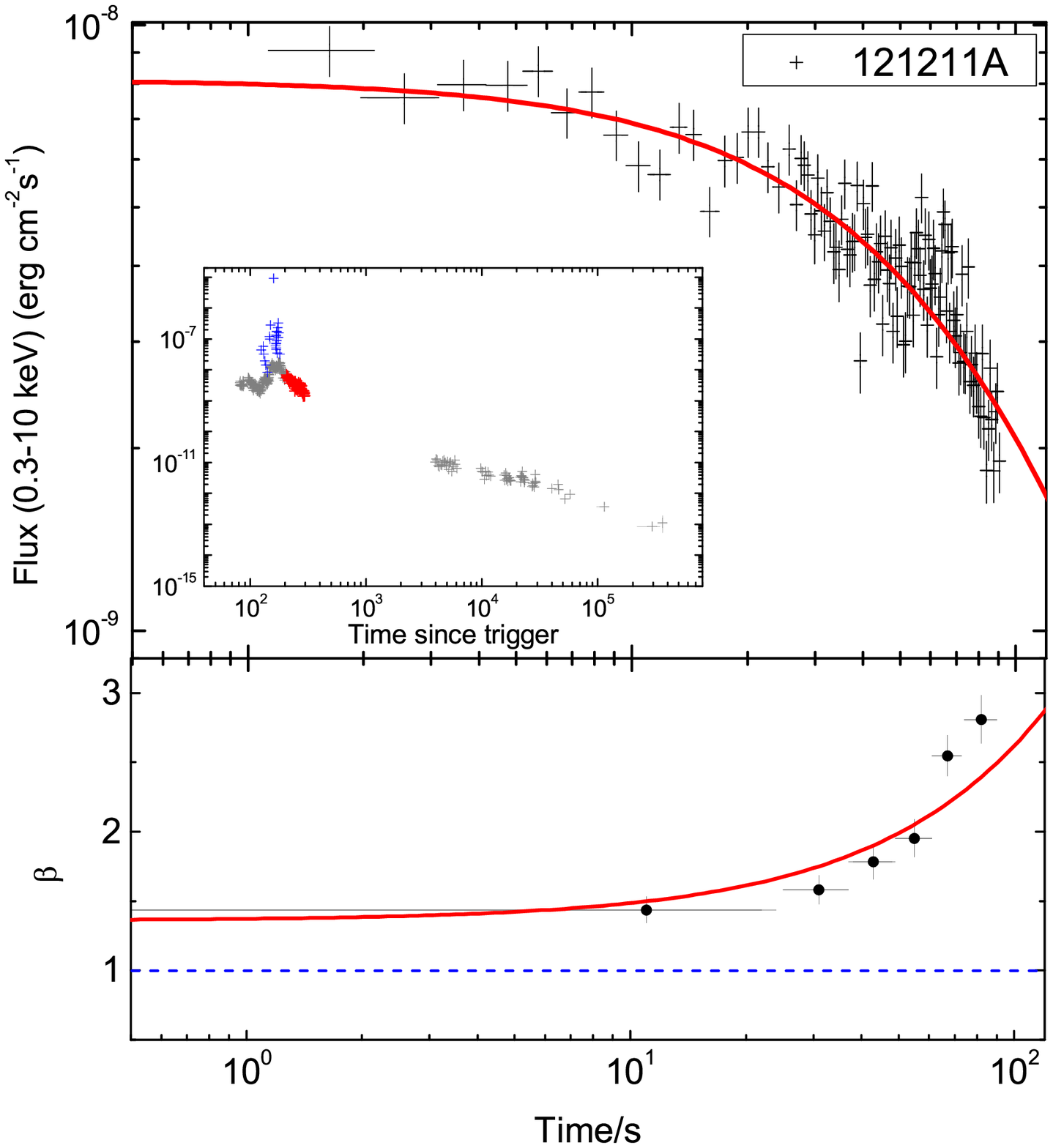}%
\includegraphics[angle=0,scale=0.350,width=0.5\textwidth,height=0.25\textheight]{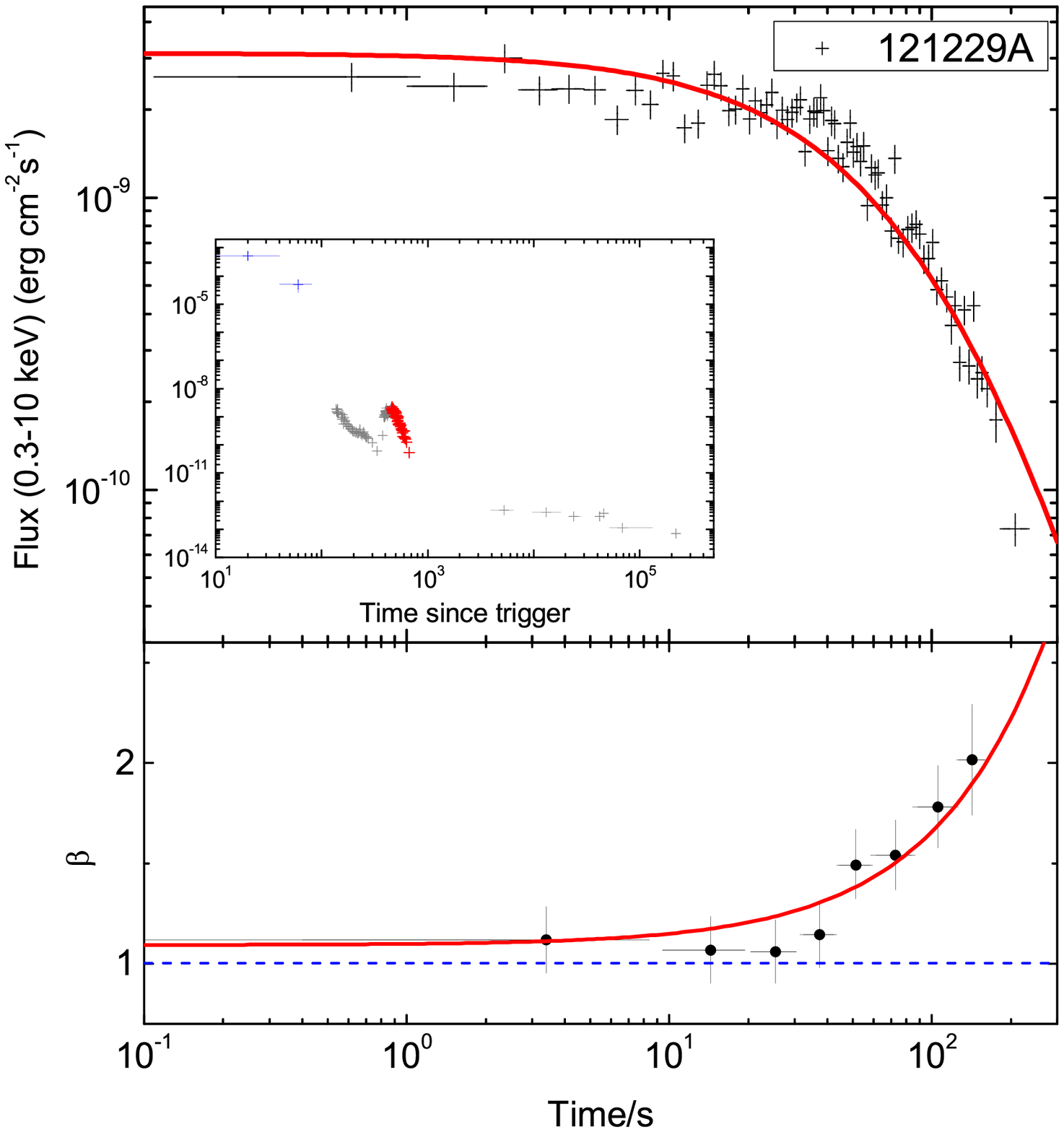}
\includegraphics[angle=0,scale=0.350,width=0.5\textwidth,height=0.25\textheight]{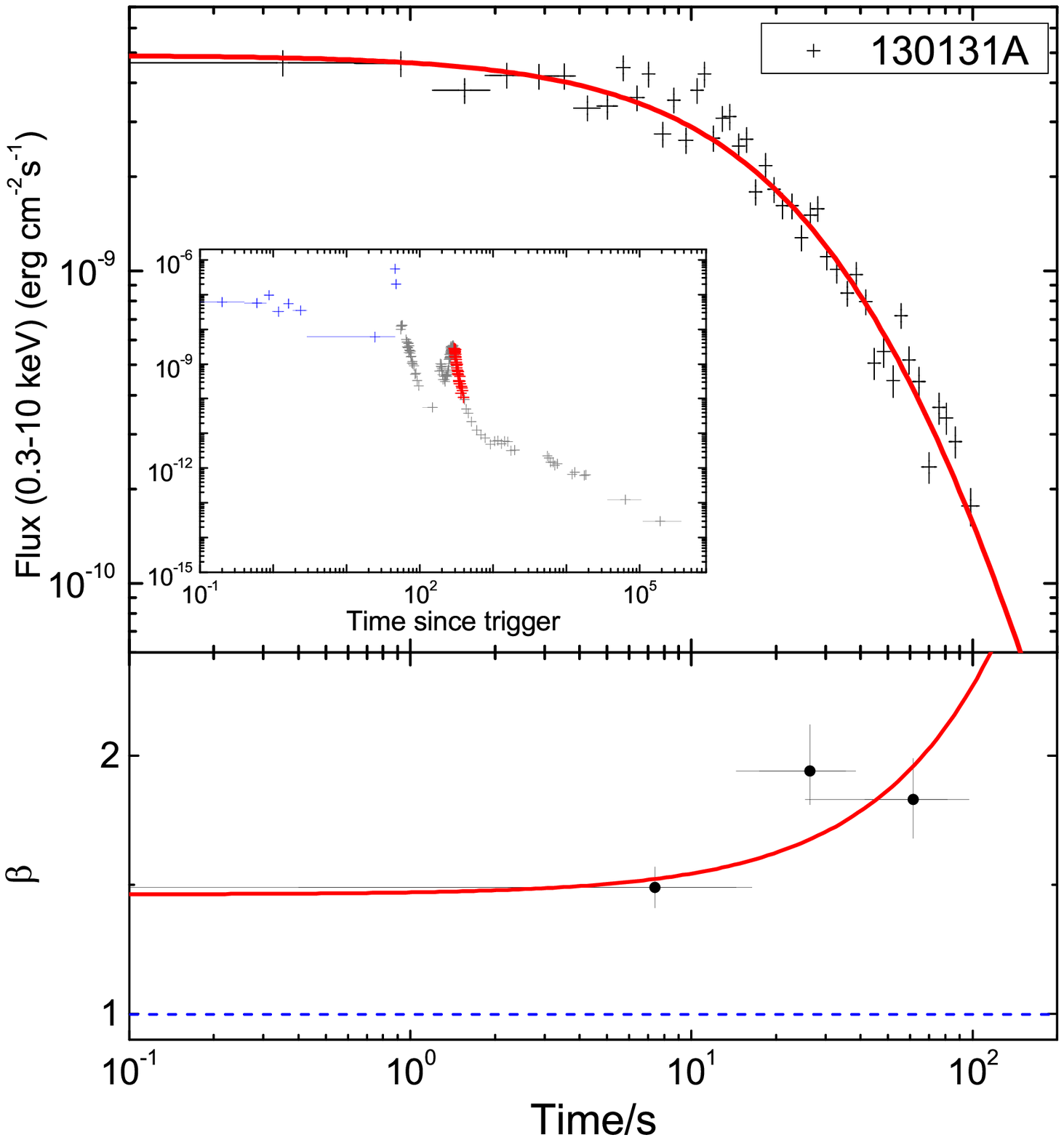}%
\includegraphics[angle=0,scale=0.350,width=0.5\textwidth,height=0.25\textheight]{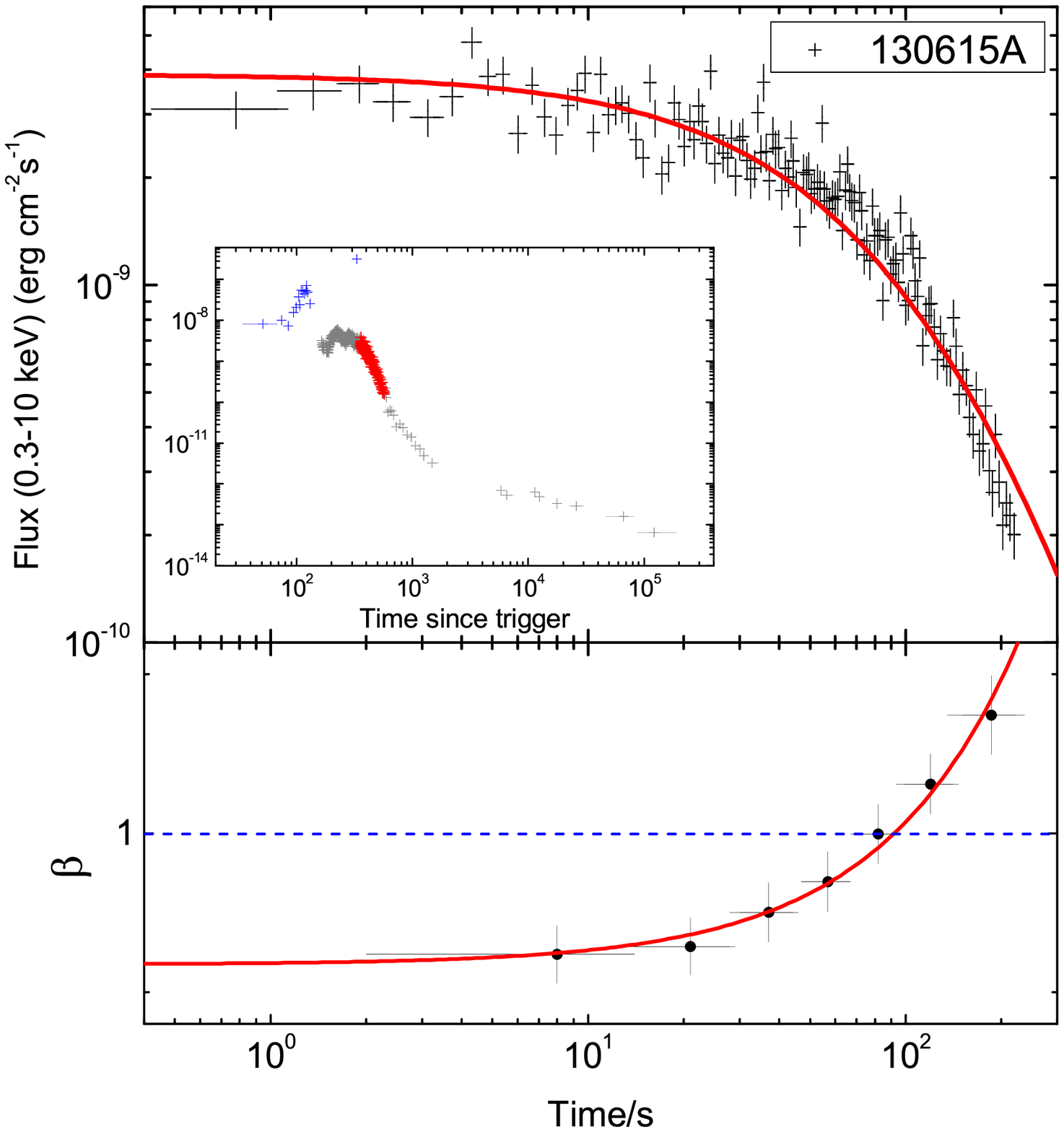}
\includegraphics[angle=0,scale=0.350,width=0.5\textwidth,height=0.25\textheight]{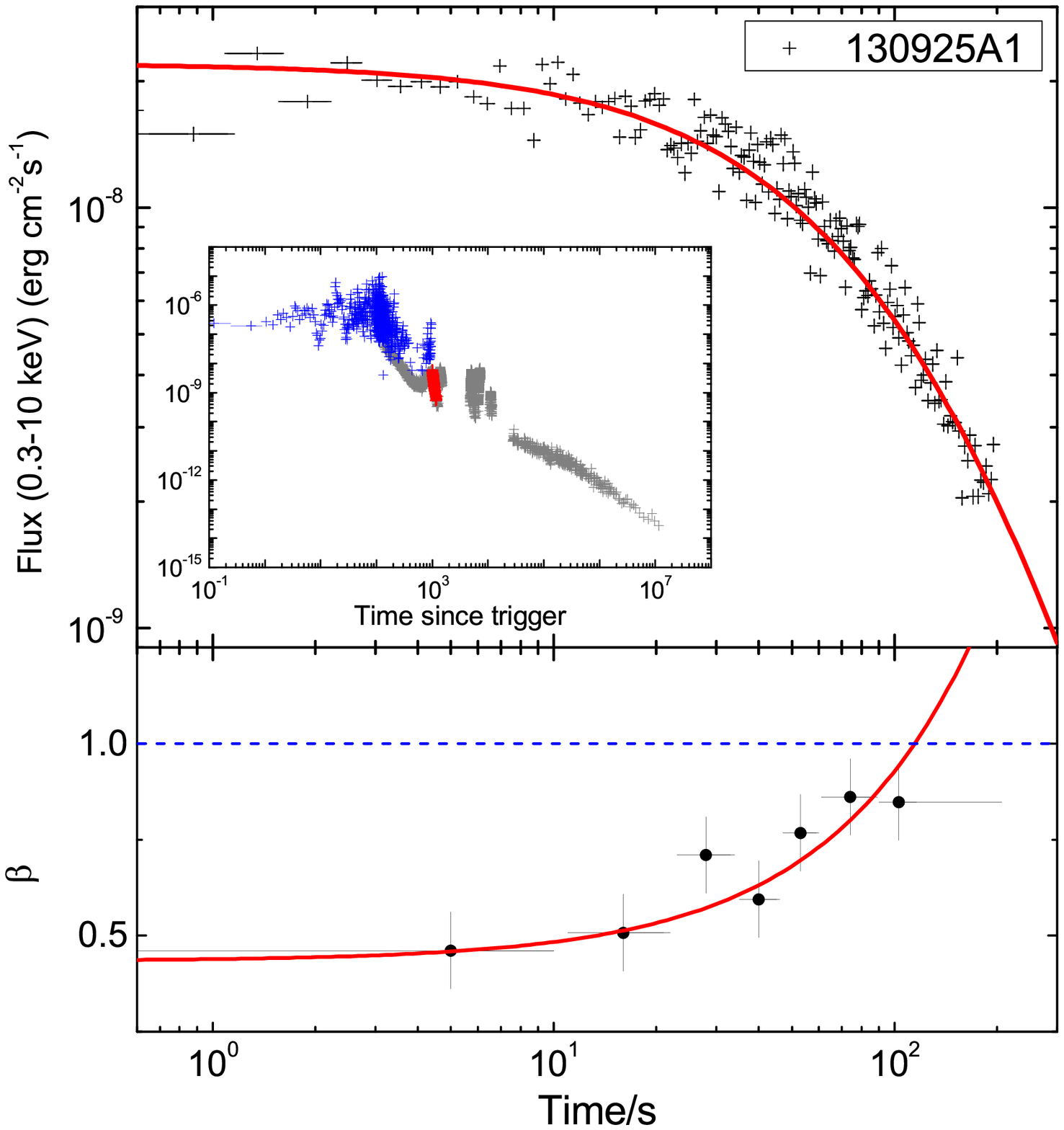}%
\includegraphics[angle=0,scale=0.350,width=0.5\textwidth,height=0.25\textheight]{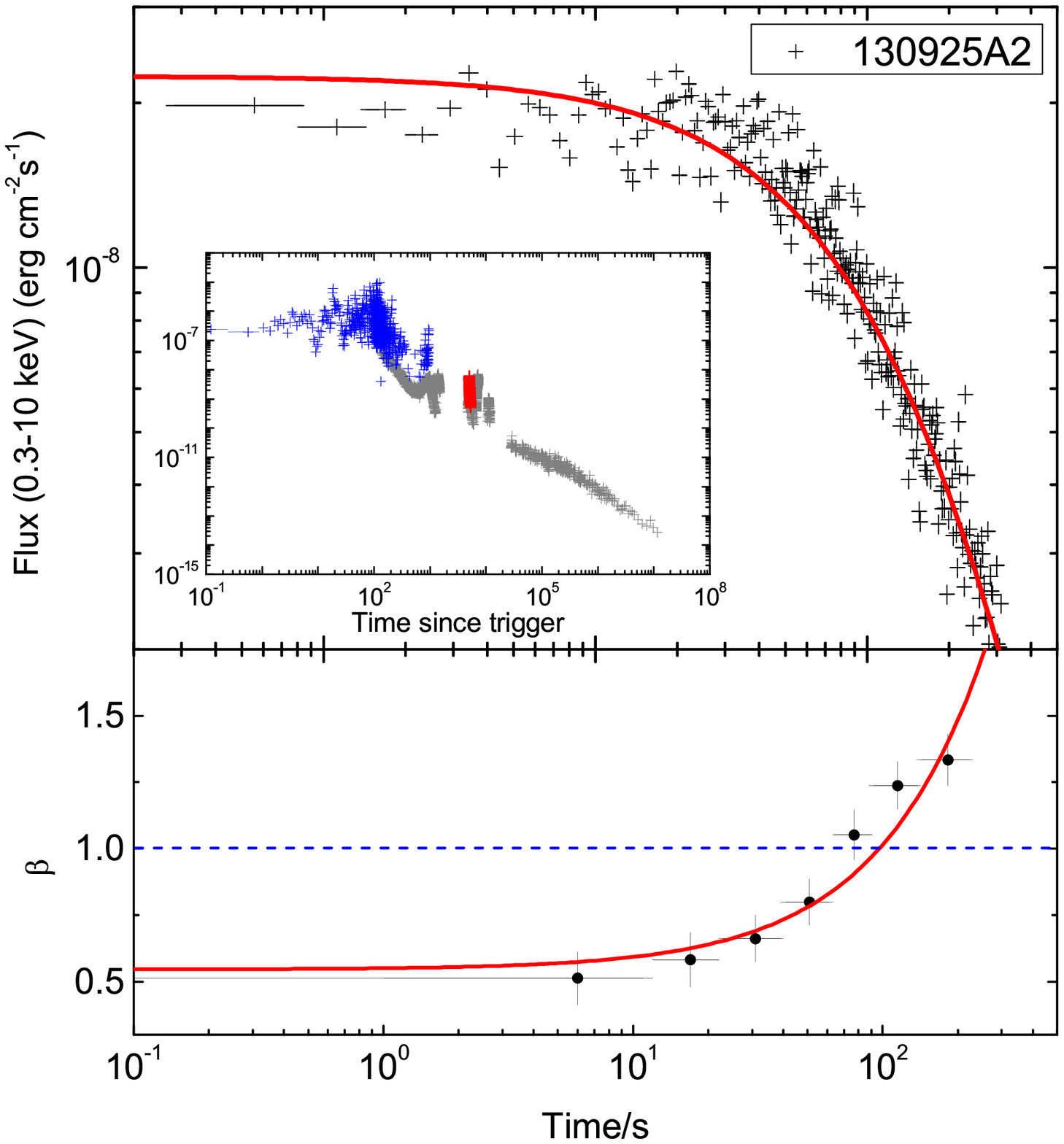}
\includegraphics[angle=0,scale=0.350,width=0.5\textwidth,height=0.25\textheight]{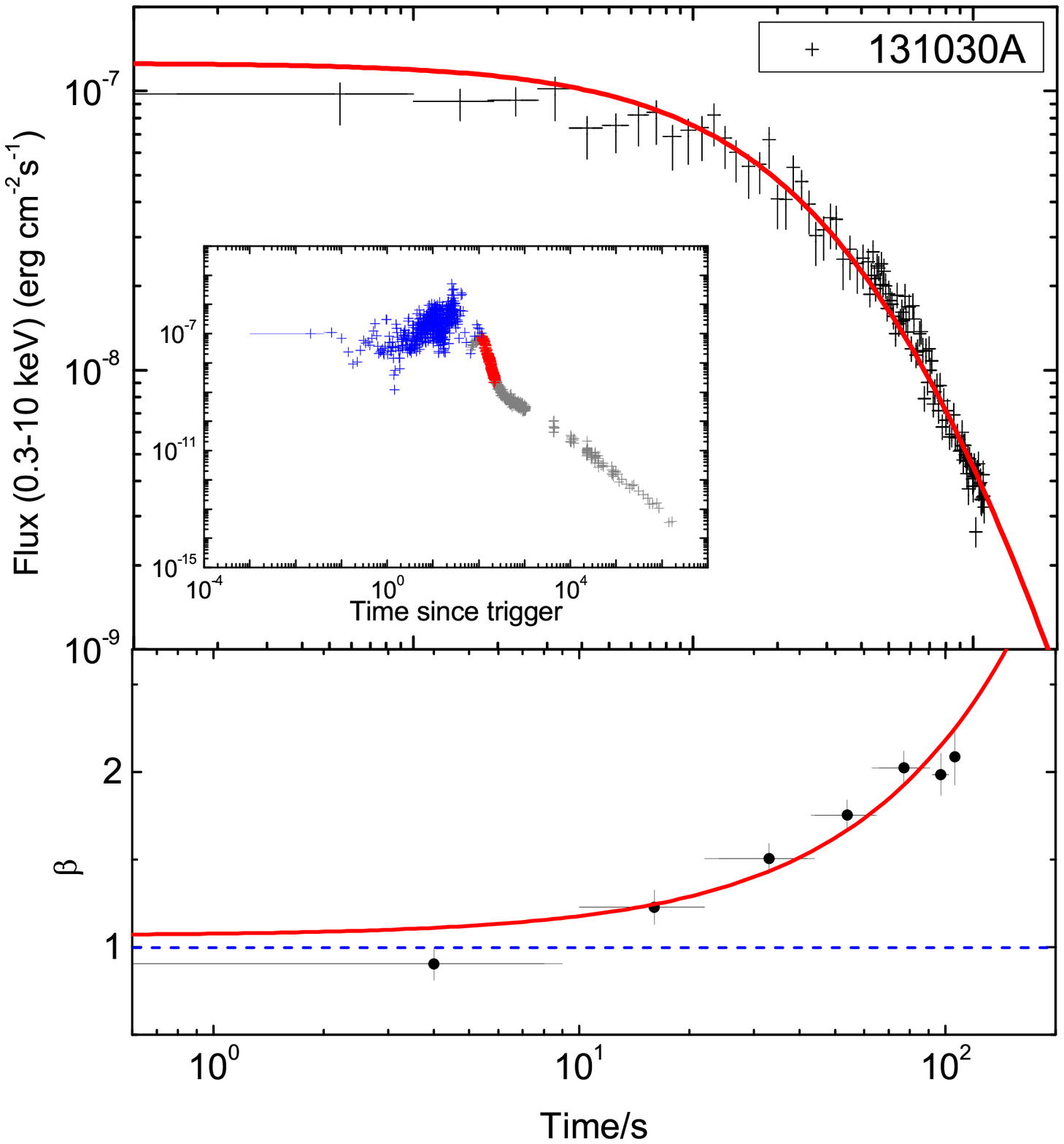}%
\includegraphics[angle=0,scale=0.350,width=0.5\textwidth,height=0.25\textheight]{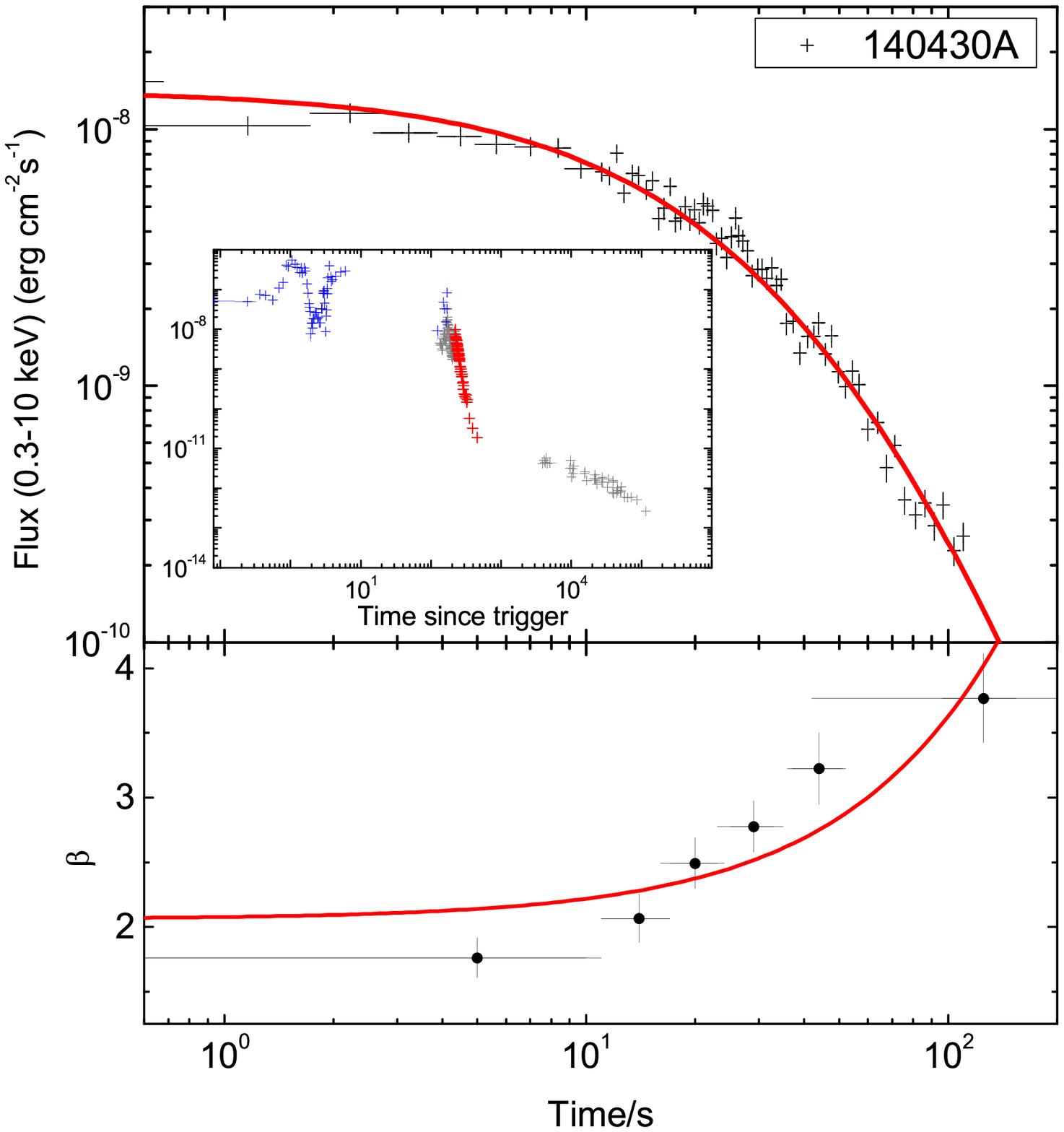}

\caption{(Continued)}
\end{figure*}
 \addtocounter{figure}{-1}
\begin{figure*}
\includegraphics[angle=0,scale=0.350,width=0.5\textwidth,height=0.25\textheight]{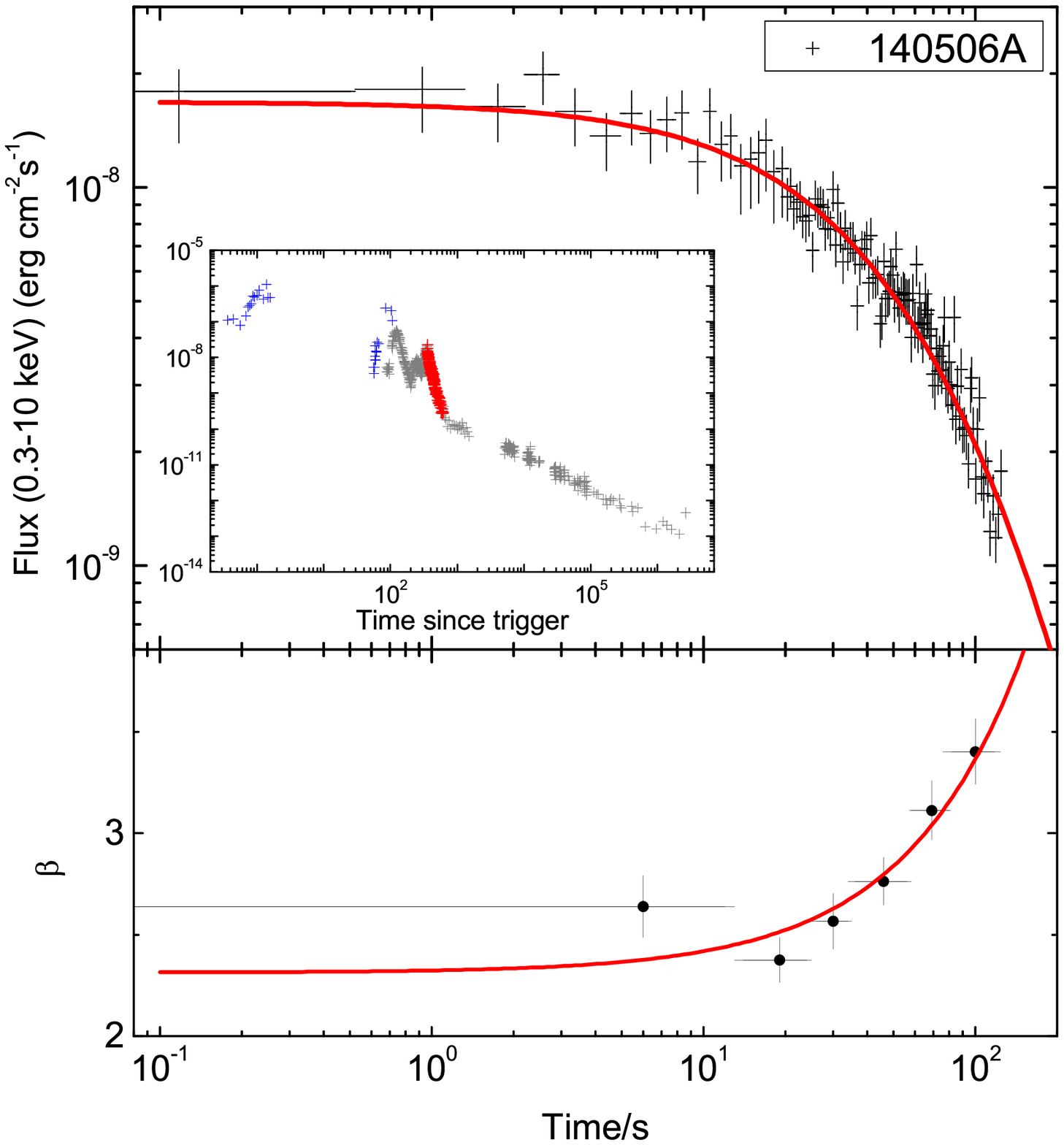}%
\includegraphics[angle=0,scale=0.350,width=0.5\textwidth,height=0.25\textheight]{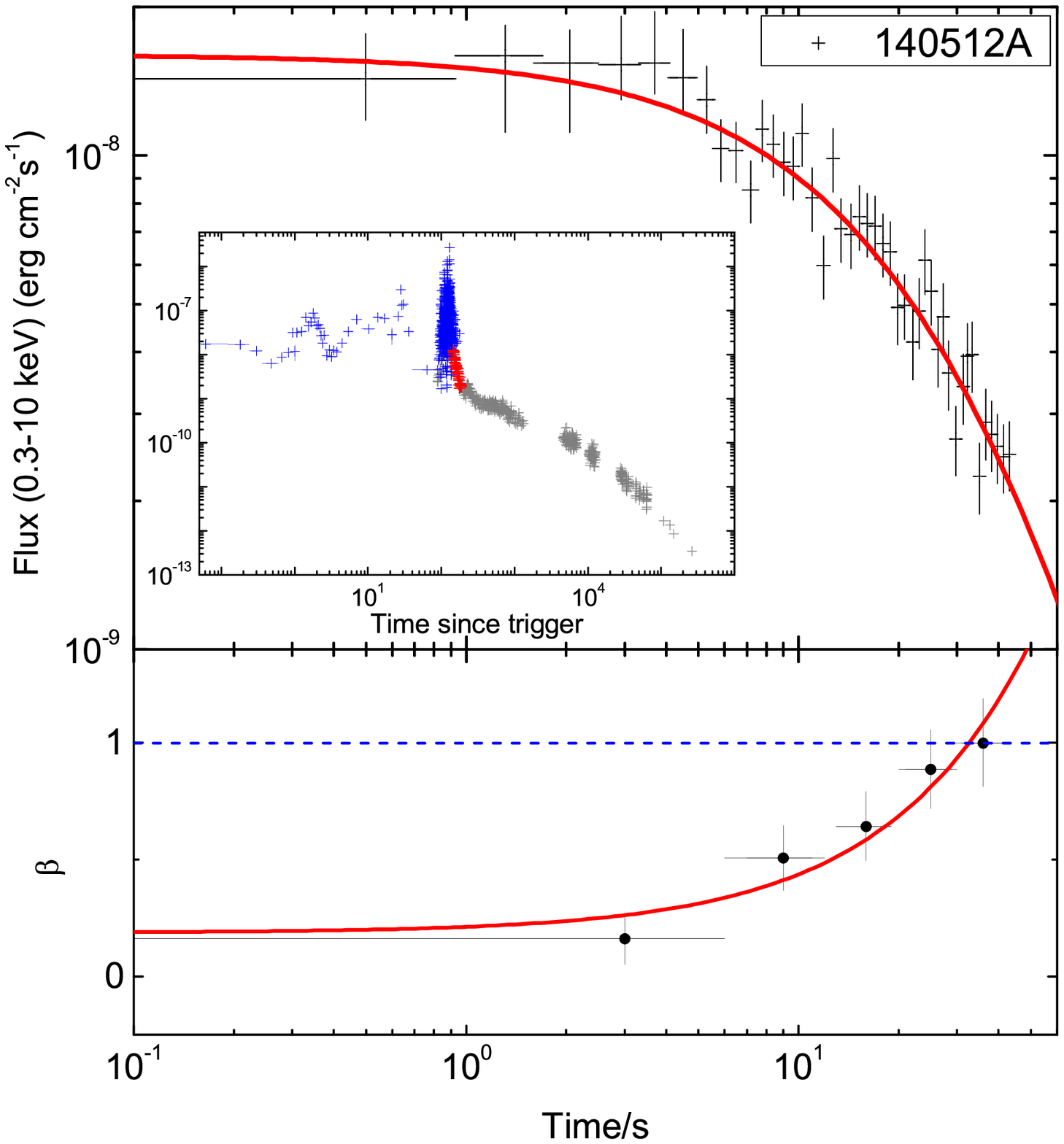}
\includegraphics[angle=0,scale=0.350,width=0.5\textwidth,height=0.25\textheight]{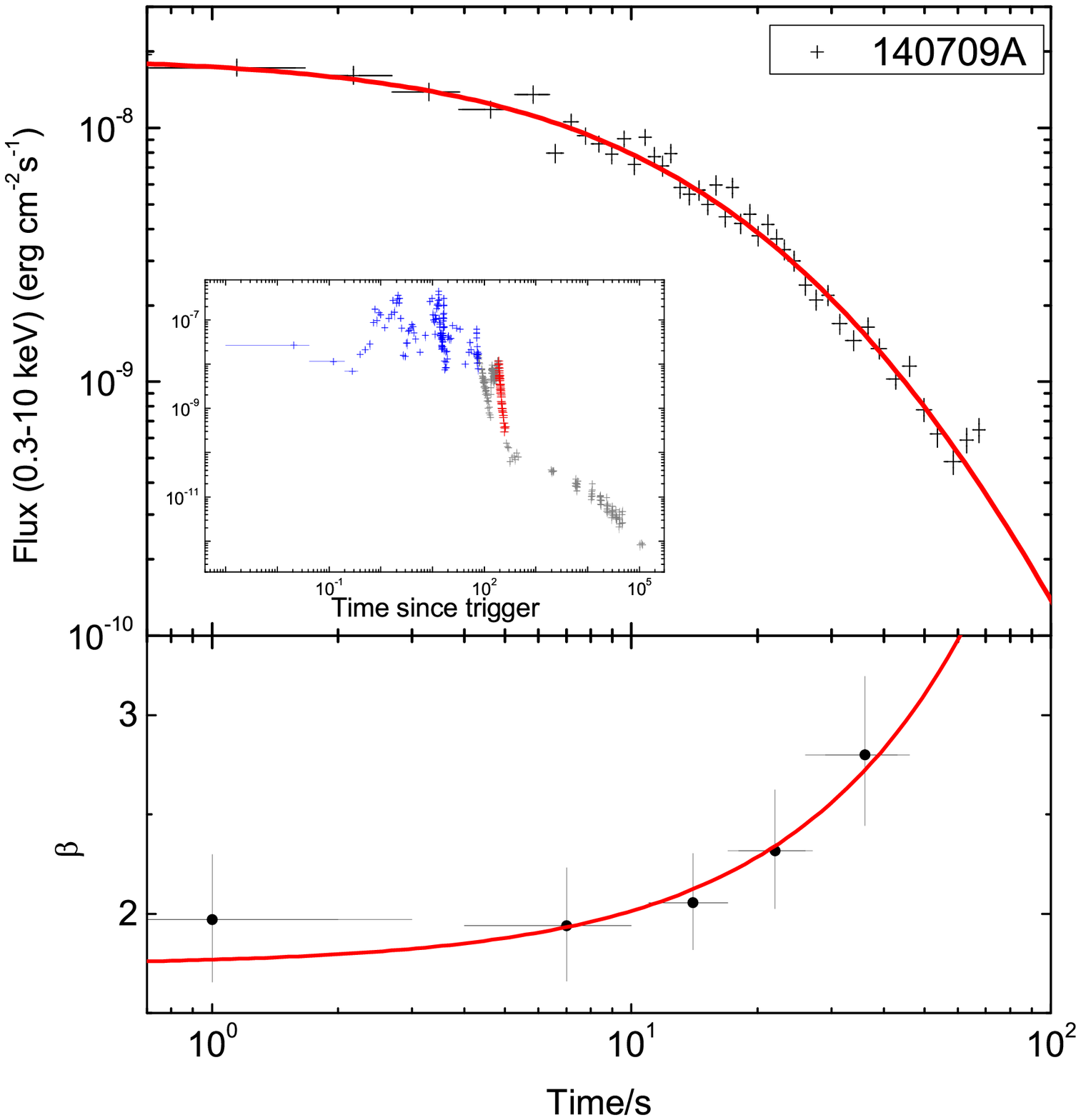}%
\includegraphics[angle=0,scale=0.350,width=0.5\textwidth,height=0.25\textheight]{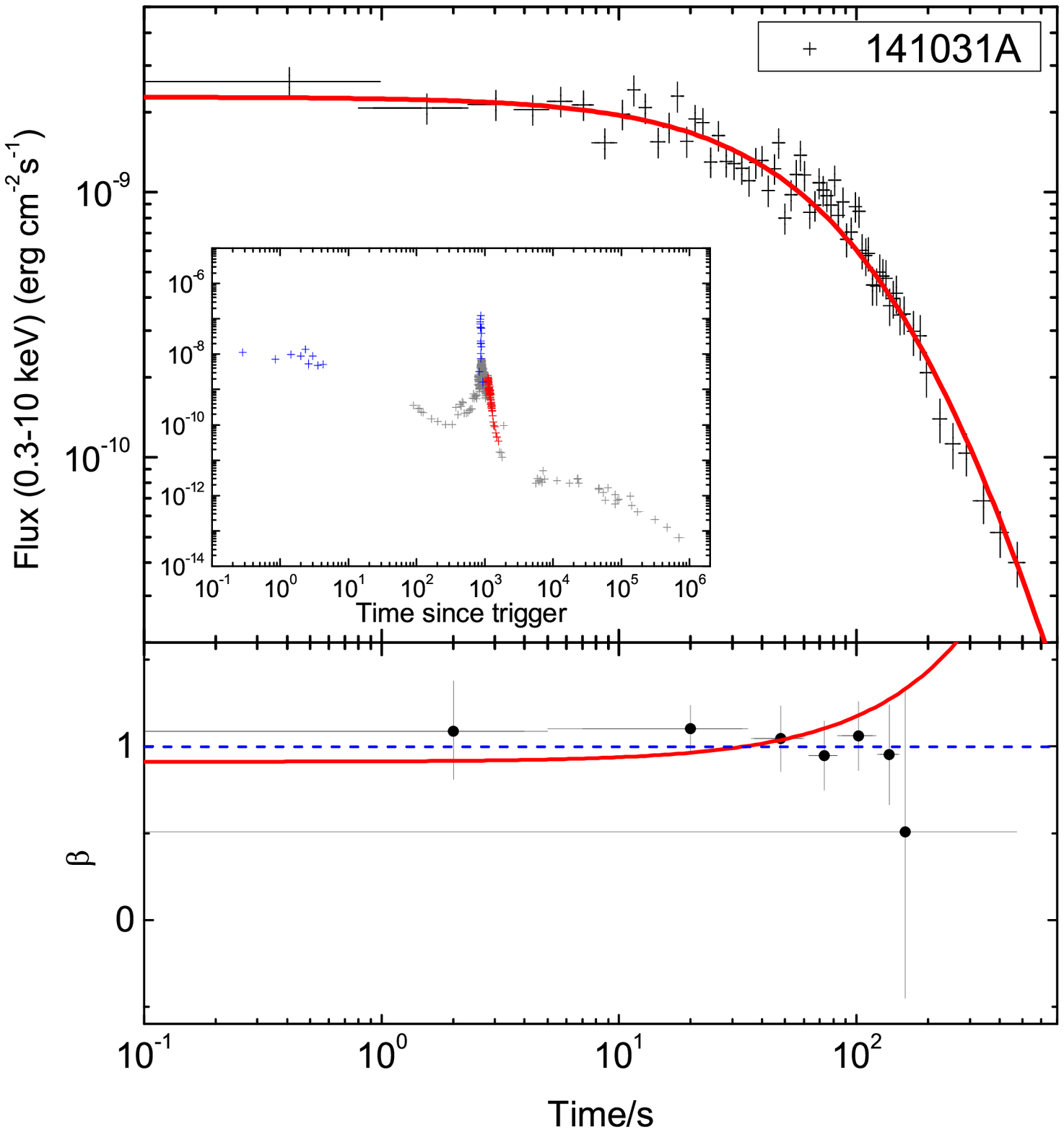}
\includegraphics[angle=0,scale=0.350,width=0.5\textwidth,height=0.25\textheight]{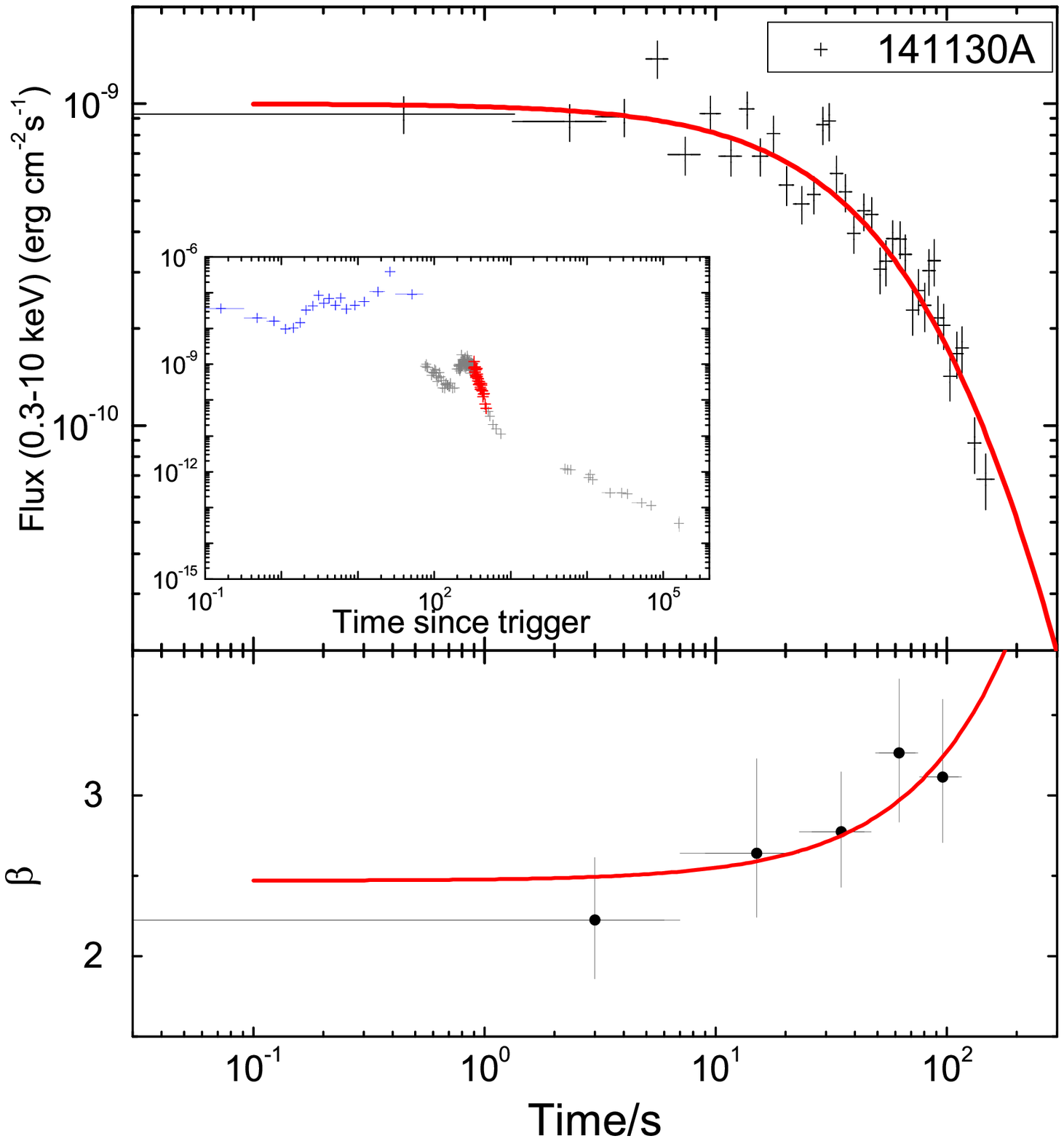}%

\caption{(Continued)}
\end{figure*}

\begin{figure}
\plotone{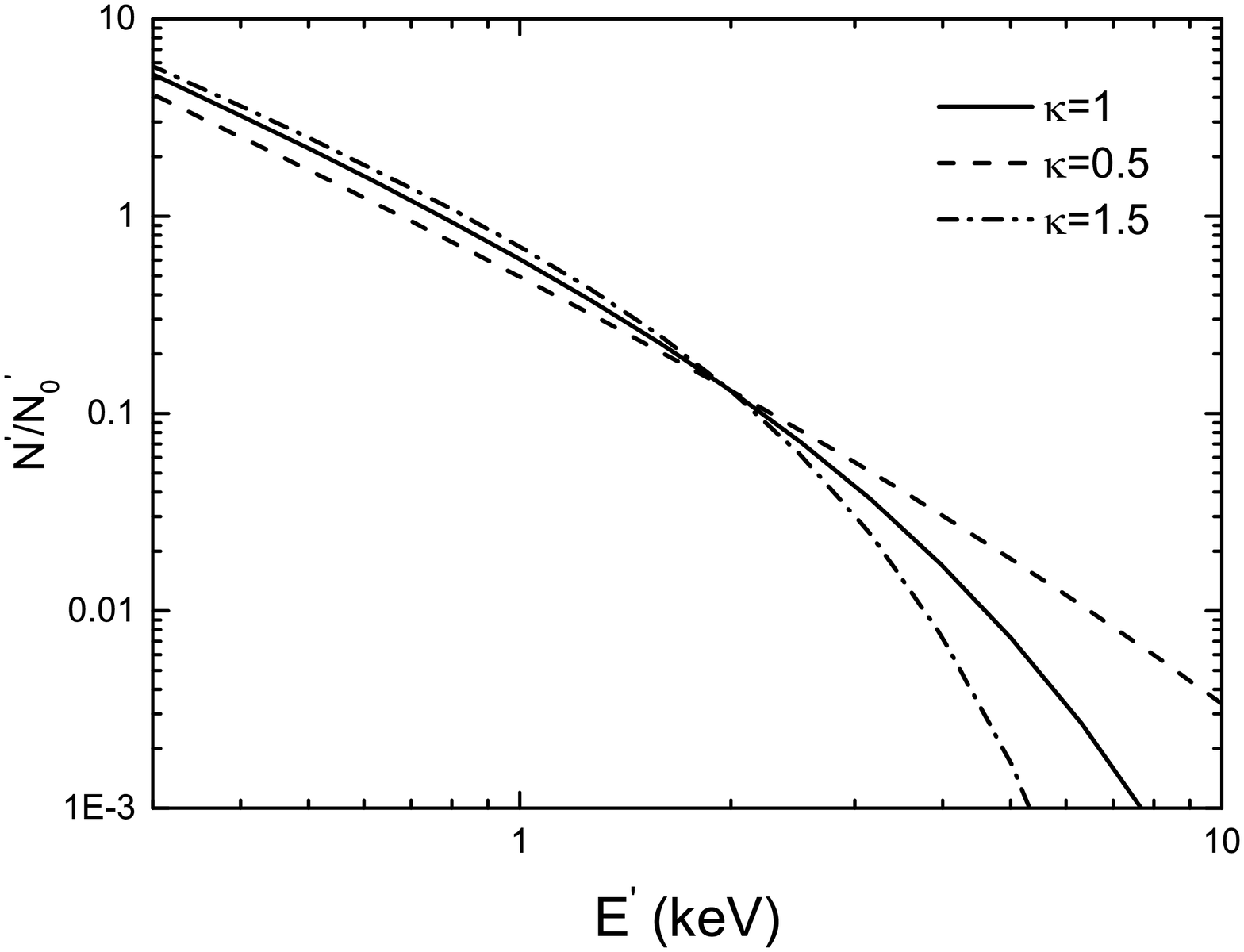}
\caption{General Cut-off power spectral model (Eq. 4) in different deepness parameter $\kappa$.}
\end{figure}

\begin{figure}
\plotone{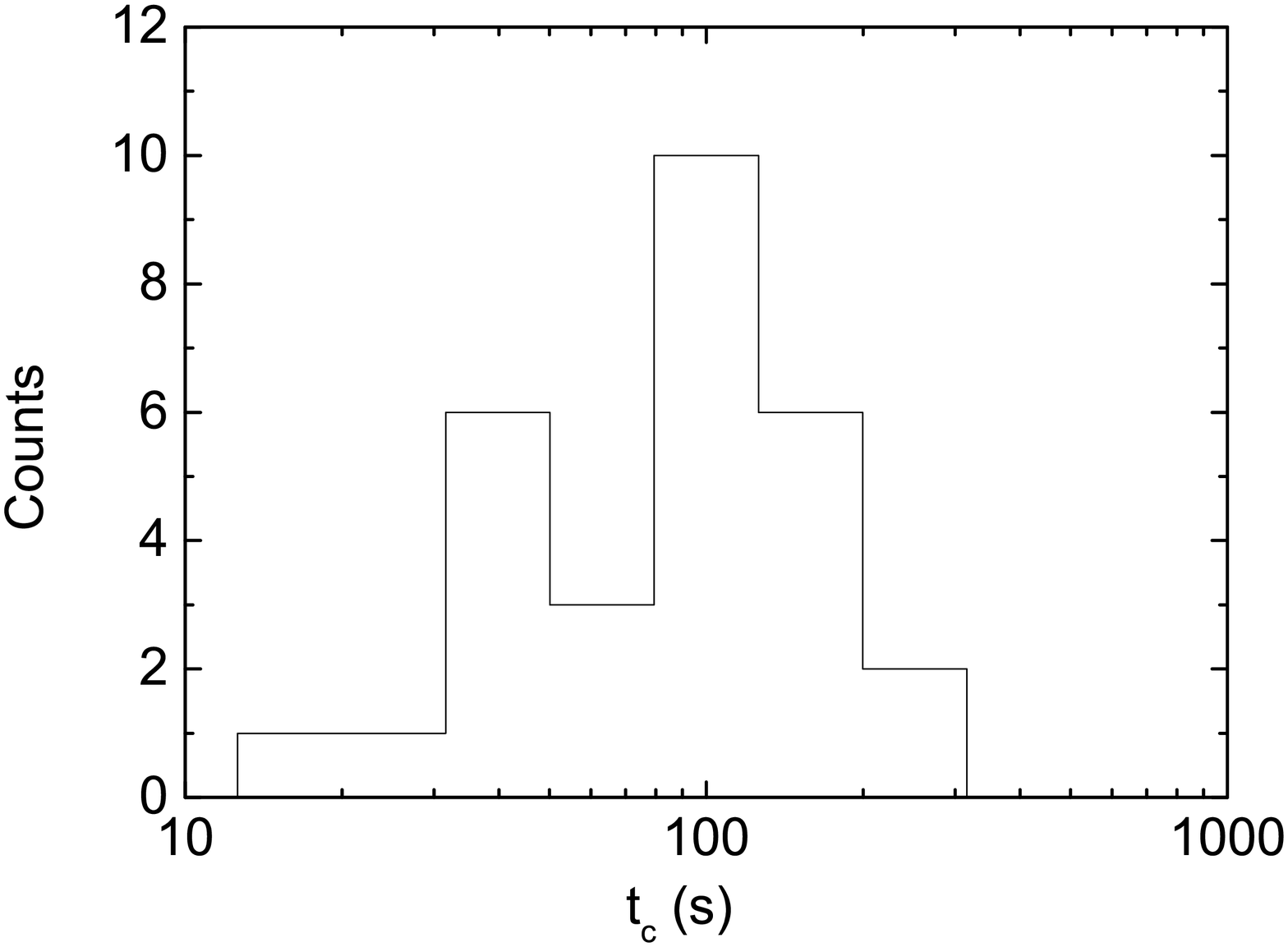}
\caption{Distributions of $t_{c}$ derived from our fits.}
\end{figure}

\begin{figure}
\plotone{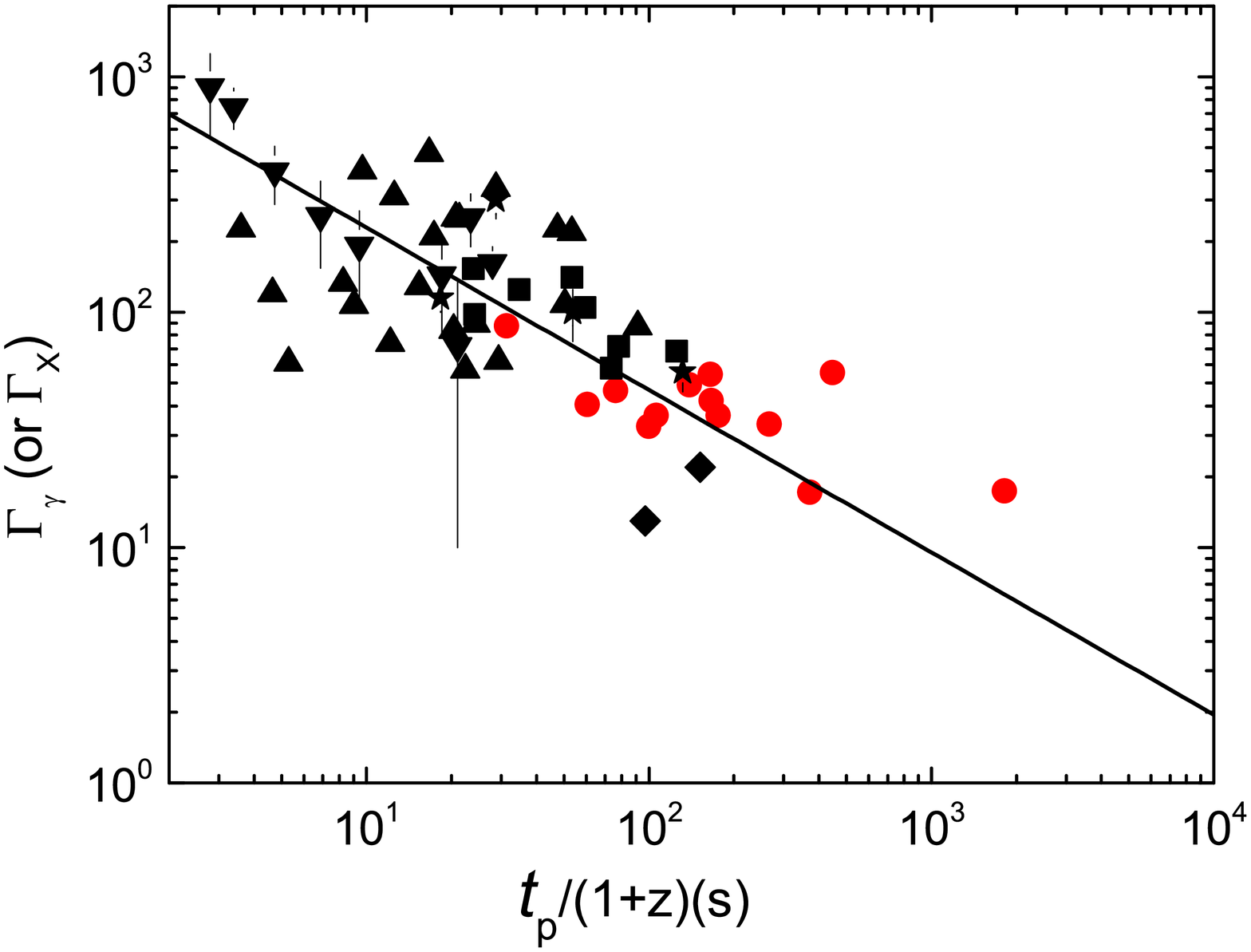}
\caption{Lorentz factor ($\Gamma_\gamma$ or $\Gamma_{\rm X}$) as a function of $t_{p}$ for both the prompt gamma-rays and X-ray flares. The red dots are the X-ray flares in our sample.
The symbols of ``$\blacktriangle$'', ``$\blacksquare$'', ``$\bigstar$'', ``$\blacktriangledown$'', and ``$\blacklozenge$''
represent the data from Liang et al.(2013), Peng et al.(2014), Troja et al.(2014) (assuming the variation timescale of 0.1s), Tang et al.(2014), and Fan et al.(2005), respectively. The line is the regression line with the Spearman correlation analysis method.}
\end{figure}

\begin{figure}
\plotone{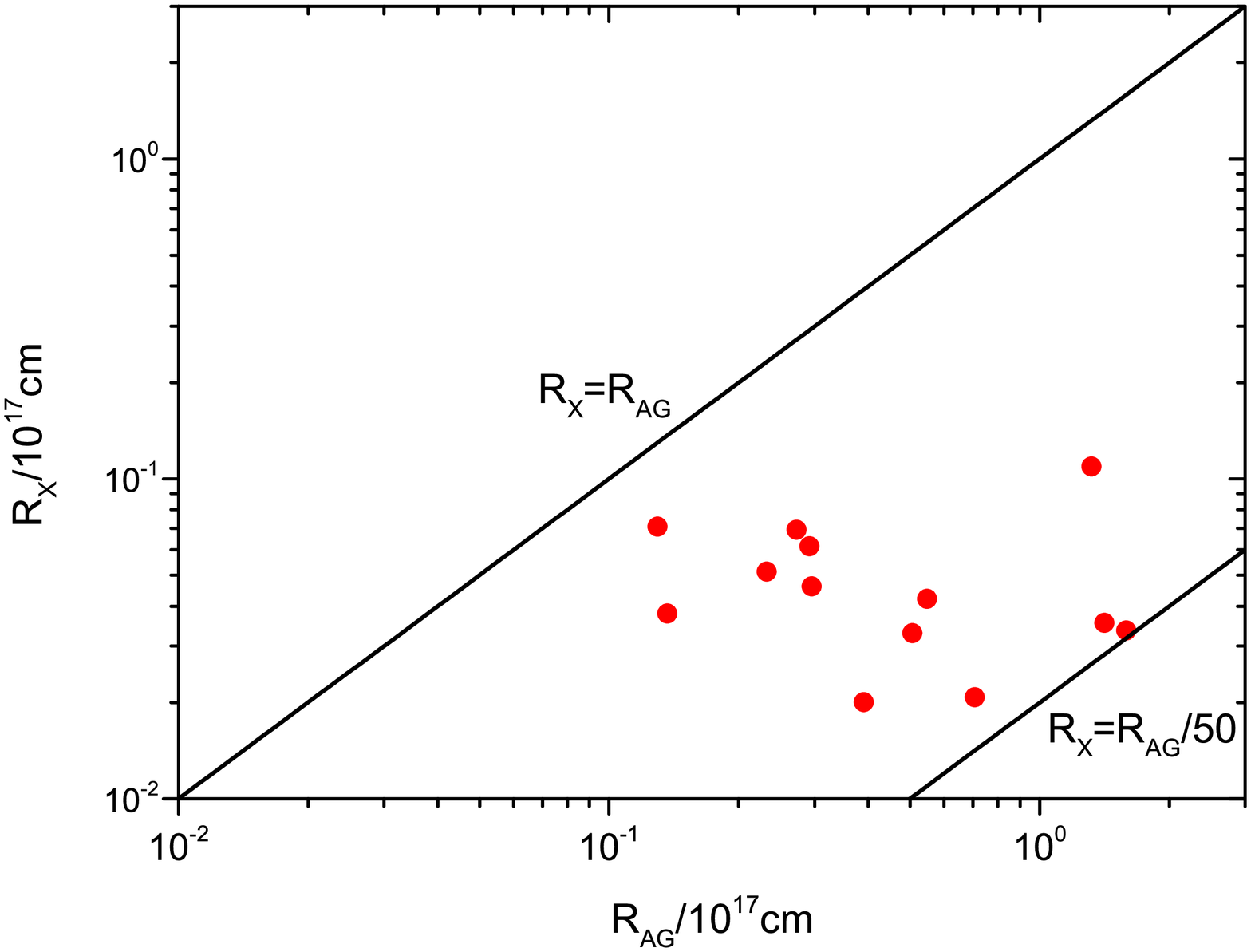}
\caption{Comparison between the radii of the X-ray flare fireballs ($R_{\rm X}$) and the radii of the afterglow fireballs at the peak time of the flares ($R_{\rm AG}$). Lines of $R_{\rm X}=R_{\rm AG}$ and $R_{\rm X}=R_{\rm AG}/50$ are also marked.}
\end{figure}

\clearpage
\begin{deluxetable}{lccccc}
\footnotesize
\tablewidth{0pc}
\tablecaption{Results of time-resolved spectral analysis for the steep decay phases of the X-ray flares in our sample}
\tablehead{\colhead{GRB} & \colhead{interval}&\colhead {mean time}& \colhead {$\beta$} & \colhead{$N_{\rm H} ^{\rm host}$} & \colhead{C\_stat/bin\tablenotemark} \\
\colhead{} & \colhead{(s)}&\colhead {(s)}& \colhead {}  & \colhead{($10^{22}$ cm$^{-2}$)} & \colhead{}
}
\startdata
050502B	&	714	-	767	&	760	&$	1.14 	_{	-0.09 	}^{+	0.09 	}$&	0.03 	&	878/859	\\
--	&	768	-	792	&	780	&$	1.17 	_{	-0.07 	}^{+	0.08 	}$&	--	&	--	\\
--	&	793	-	825	&	808	&$	1.32 	_{	-0.08 	}^{+	0.08 	}$&	--	&	--	\\
--	&	827	-	887	&	853	&$	1.53 	_{	-0.08 	}^{+	0.08 	}$&	--	&	--	\\
--	&	892	-	1021	&	945	&$	1.66 	_{	-0.09 	}^{+	0.10 	}$&	--	&	--	\\
--	&	1031	-	1130	&	1062	&$	1.96 	_{	-0.21 	}^{+	0.22 	}$&	--	&	--	\\
060111A	&	283	-	294	&	288	&$	0.89 	_{	-0.08 	}^{+	0.09 	}$&	0.13 	&	810/743	\\
--	&	295	-	319	&	308	&$	1.12 	_{	-0.10 	}^{+	0.10 	}$&	--	&	--	\\
--	&	320	-	331	&	326	&$	0.83 	_{	-0.10 	}^{+	0.10 	}$&	--	&	--	\\
--	&	332	-	347	&	339	&$	1.06 	_{	-0.10 	}^{+	0.10 	}$&	--	&	--	\\
--	&	347	-	370	&	358	&$	1.10 	_{	-0.11 	}^{+	0.11 	}$&	--	&	--	\\
--	&	371	-	410	&	387	&$	1.05 	_{	-0.10 	}^{+	0.11 	}$&	--	&	--	\\
--	&	413	-	483	&	443	&$	1.19 	_{	-0.12 	}^{+	0.12 	}$&	--	&	--	\\
060124	&	761	-	771	&	766	&$	2.00 	_{	-0.08 	}^{+	0.08 	}$&	0.13 	&	630/623	\\
--	&	772	-	782	&	777	&$	2.12 	_{	-0.09 	}^{+	0.10 	}$&	--	&	--	\\
--	&	782	-	797	&	789	&$	2.32 	_{	-0.10 	}^{+	0.10 	}$&	--	&	--	\\
--	&	798	-	813	&	805	&$	2.38 	_{	-0.10 	}^{+	0.10 	}$&	--	&	--	\\
--	&	814	-	836	&	824	&$	2.42 	_{	-0.10 	}^{+	0.11 	}$&	--	&	--	\\
060904B	&	172	-	184	&	178	&$	0.93 	_{	-0.09 	}^{+	0.09 	}$&	0.49 	&	692/645	\\
--	&	187	-	220	&	196	&$	1.33 	_{	-0.10 	}^{+	0.10 	}$&	--	&	--	\\
--	&	207	-	252	&	212	&$	1.66 	_{	-0.16 	}^{+	0.17 	}$&	--	&	--	\\
--	&	219	-	278	&	225	&$	1.86 	_{	-0.10 	}^{+	0.10 	}$&	--	&	--	\\
--	&	233	-	308	&	240	&$	2.22 	_{	-0.12 	}^{+	0.12 	}$&	--	&	--	\\
--	&	249	-	340	&	256	&$	2.52 	_{	-0.14 	}^{+	0.15 	}$&	--	&	--	\\
--	&	268	-	396	&	284	&$	2.98 	_{	-0.17 	}^{+	0.18 	}$&	--	&	--	\\
060929	&	527	-	542	&	535	&$	0.46 	_{	-0.10 	}^{+	0.10 	}$&	0.11 	&	500/470	\\
--	&	543	-	557	&	550	&$	0.77 	_{	-0.15 	}^{+	0.16 	}$&	--	&	--	\\
--	&	557	-	574	&	565	&$	0.80 	_{	-0.11 	}^{+	0.11 	}$&	--	&	--	\\
--	&	575	-	600	&	587	&$	0.97 	_{	-0.11 	}^{+	0.11 	}$&	--	&	--	\\
--	&	602	-	640	&	620	&$	1.24 	_{	-0.12 	}^{+	0.12 	}$&	--	&	--	\\
--	&	642	-	709	&	668	&$	1.40 	_{	-0.13 	}^{+	0.13 	}$&	--	&	--	\\
070520B	&	180	-	184	&	182	&$	1.01 	_{	-0.14 	}^{+	0.15 	}$&	0.20 	&	563/453	\\
--	&	185	-	197	&	190	&$	1.47 	_{	-0.17 	}^{+	0.17 	}$&	--	&	--	\\
--	&	197	-	214	&	205	&$	1.50 	_{	-0.10 	}^{+	0.10 	}$&	--	&	--	\\
--	&	216	-	239	&	227	&$	1.77 	_{	-0.11 	}^{+	0.11 	}$&	--	&	--	\\
--	&	239	-	269	&	253	&$	2.00 	_{	-0.12 	}^{+	0.13 	}$&	--	&	--	\\
--	&	270	-	340	&	298	&$	2.05 	_{	-0.14 	}^{+	0.14 	}$&	--	&	--	\\
070704	&	338	-	350	&	344	&$	0.93 	_{	-0.13 	}^{+	0.13 	}$&	0.32 	&	977/924	\\
--	&	351	-	365	&	358	&$	0.96 	_{	-0.13 	}^{+	0.13 	}$&	--	&	--	\\
--	&	366	-	386	&	376	&$	1.04 	_{	-0.13 	}^{+	0.13 	}$&	--	&	--	\\
--	&	387	-	420	&	401	&$	1.26 	_{	-0.13 	}^{+	0.13 	}$&	--	&	--	\\
--	&	428	-	523	&	468	&$	1.55 	_{	-0.18 	}^{+	0.19 	}$&	--	&	--	\\
080928	&	355	-	361	&	358	&$	0.39 	_{	-0.13 	}^{+	0.14 	}$&	0.38 	&	259/257	\\
--	&	361	-	368	&	364	&$	0.69 	_{	-0.15 	}^{+	0.15 	}$&	--	&	--	\\
--	&	369	-	377	&	373	&$	0.78 	_{	-0.15 	}^{+	0.15 	}$&	--	&	--	\\
--	&	378	-	391	&	384	&$	0.86 	_{	-0.16 	}^{+	0.17 	}$&	--	&	--	\\
--	&	392	-	412	&	402	&$	0.97 	_{	-0.17 	}^{+	0.18 	}$&	--	&	--	\\
091130B	&	103	-	112	&	107	&$	0.91 	_{	-0.12 	}^{+	0.12 	}$&	0.31 	&	657/665	\\
--	&	112	-	122	&	117	&$	1.40 	_{	-0.19 	}^{+	0.20 	}$&	--	&	--	\\
--	&	123	-	132	&	128	&$	1.61 	_{	-0.16 	}^{+	0.17 	}$&	--	&	--	\\
--	&	134	-	146	&	139	&$	1.70 	_{	-0.16 	}^{+	0.17 	}$&	--	&	--	\\
--	&	147	-	161	&	154	&$	1.85 	_{	-0.20 	}^{+	0.21 	}$&	--	&	--	\\
--	&	163	-	177	&	169	&$	2.30 	_{	-0.25 	}^{+	0.26 	}$&	--	&	--	\\
100619A	&	949	-	968	&	959	&$	0.73 	_{	-0.10 	}^{+	0.10 	}$&	0.47 	&	818/772	\\
--	&	969	-	991	&	980	&$	1.01 	_{	-0.15 	}^{+	0.16 	}$&	--	&	--	\\
--	&	992	-	1019	&	1005	&$	1.12 	_{	-0.10 	}^{+	0.10 	}$&	--	&	--	\\
--	&	1020	-	1052	&	1036	&$	1.43 	_{	-0.11 	}^{+	0.11 	}$&	--	&	--	\\
--	&	1053	-	1097	&	1073	&$	1.44 	_{	-0.11 	}^{+	0.11 	}$&	--	&	--	\\
--	&	1100	-	1191	&	1137	&$	1.64 	_{	-0.12 	}^{+	0.12 	}$&	--	&	--	\\
--	&	1195	-	1286	&	1234	&$	2.06 	_{	-0.18 	}^{+	0.18 	}$&	--	&	--	\\
100704A	&	196	-	228	&	208	&$	1.37 	_{	-0.08 	}^{+	0.07 	}$&	0.31 	&	1064/820	 \\
--	&	228	-	242	&	234	&$	2.10 	_{	-0.12 	}^{+	0.13 	}$&	--	&	--	\\
--	&	242	-	254	&	248	&$	2.43 	_{	-0.16 	}^{+	0.17 	}$&	--	&	--	\\
--	&	255	-	264	&	259	&$	2.60 	_{	-0.19 	}^{+	0.20 	}$&	--	&	--	\\
--	&	265	-	275	&	270	&$	2.96 	_{	-0.23 	}^{+	0.23 	}$&	--	&	--	\\
--	&	270	-	330	&	300	&$	3.24 	_{	-0.13 	}^{+	0.13 	}$&	--	&	--	\\
100802A	&	507	-	538	&	522	&$	0.81 	_{	-0.07 	}^{+	0.07 	}$&	0.03 	&	995/826	\\
--	&	541	-	564	&	552	&$	1.04 	_{	-0.13 	}^{+	0.13 	}$&	--	&	--	\\
--	&	567	-	594	&	581	&$	1.03 	_{	-0.09 	}^{+	0.09 	}$&	--	&	--	\\
--	&	598	-	635	&	615	&$	1.16 	_{	-0.09 	}^{+	0.09 	}$&	--	&	--	\\
--	&	640	-	707	&	668	&$	1.22 	_{	-0.09 	}^{+	0.10 	}$&	--	&	--	\\
--	&	719	-	818	&	784	&$	1.41 	_{	-0.12 	}^{+	0.14 	}$&	--	&	--	\\
100902A	&	418	-	431	&	424	&$	0.95 	_{	-0.07 	}^{+	0.05 	}$&	--	&	--	\\
--	&	432	-	454	&	442	&$	1.86 	_{	-0.10 	}^{+	0.10 	}$&	--	&	--	\\
--	&	457	-	473	&	464	&$	2.31 	_{	-0.14 	}^{+	0.13 	}$&	--	&	--	\\
--	&	473	-	482	&	477	&$	2.49 	_{	-0.15 	}^{+	0.14 	}$&	--	&	--	\\
--	&	482	-	492	&	487	&$	2.65 	_{	-0.16 	}^{+	0.16 	}$&	--	&	--	\\
--	&	493	-	511	&	501	&$	2.84 	_{	-0.15 	}^{+	0.15 	}$&	--	&	--	\\
110801A	&	453	-	472	&	462	&$	1.53 	_{	-0.09 	}^{+	0.09 	}$&	--	&	1103/833	\\
--	&	473	-	486	&	479	&$	1.71 	_{	-0.13 	}^{+	0.13 	}$&	--	&	--	\\
--	&	487	-	507	&	496	&$	1.81 	_{	-0.13 	}^{+	0.13 	}$&	--	&	--	\\
110820A	&	258	-	262	&	260	&$	-0.08 	_{	-0.21 	}^{+	0.21 	}$&	--	&	181/157	\\
--	&	262	-	267	&	265	&$	0.32 	_{	-0.36 	}^{+	0.42 	}$&	--	&	--	\\
--	&	267	-	271	&	269	&$	0.51 	_{	-0.20 	}^{+	0.21 	}$&	--	&	--	\\
--	&	298	-	328	&	309	&$	1.28 	_{	-0.28 	}^{+	0.30 	}$&	--	&	--	\\
121027A	&	259	-	277	&	268	&$	1.19 	_{	-0.13 	}^{+	0.14 	}$&	1.33 	&	338/322	\\
--	&	278	-	303	&	290	&$	1.58 	_{	-0.14 	}^{+	0.15 	}$&	--	&	--	\\
--	&	305	-	337	&	320	&$	1.50 	_{	-0.14 	}^{+	0.15 	}$&	--	&	--	\\
--	&	339	-	392	&	363	&$	1.56 	_{	-0.15 	}^{+	0.15 	}$&	--	&	--	\\
--	&	397	-	441	&	418	&$	1.65 	_{	-0.18 	}^{+	0.19 	}$&	--	&	--	\\
--	&	446	-	494	&	467	&$	1.44 	_{	-0.23 	}^{+	0.23 	}$&	--	&	--	\\
121211A	&	182	-	203	&	192	&$	1.44 	_{	-0.10 	}^{+	0.10 	}$&	1.20 	&	1211/818	 \\
--	&	204	-	228	&	215	&$	1.58 	_{	-0.10 	}^{+	0.11 	}$&	--	&	--	\\
--	&	229	-	241	&	235	&$	1.78 	_{	-0.13 	}^{+	0.13 	}$&	--	&	--	\\
--	&	242	-	253	&	247	&$	1.95 	_{	-0.13 	}^{+	0.14 	}$&	--	&	--	\\
--	&	253	-	265	&	259	&$	2.55 	_{	-0.15 	}^{+	0.15 	}$&	--	&	--	\\
--	&	265	-	277	&	271	&$	2.81 	_{	-0.17 	}^{+	0.18 	}$&	--	&	--	\\
121229A	&	458	-	467	&	462	&$	1.12 	_{	-0.17 	}^{+	0.17 	}$&	0.50 	&	352/309	\\
--	&	468	-	478	&	473	&$	1.07 	_{	-0.17 	}^{+	0.17 	}$&	--	&	--	\\
--	&	479	-	489	&	484	&$	1.06 	_{	-0.16 	}^{+	0.16 	}$&	--	&	--	\\
--	&	490	-	501	&	496	&$	1.14 	_{	-0.17 	}^{+	0.17 	}$&	--	&	--	\\
--	&	490	-	518	&	510	&$	1.49 	_{	-0.17 	}^{+	0.18 	}$&	--	&	--	\\
--	&	490	-	545	&	531	&$	1.54 	_{	-0.17 	}^{+	0.18 	}$&	--	&	--	\\
--	&	490	-	585	&	564	&$	1.78 	_{	-0.20 	}^{+	0.21 	}$&	--	&	--	\\
--	&	490	-	619	&	601	&$	2.01 	_{	-0.27 	}^{+	0.28 	}$&	--	&	--	\\
130131A	&	293	-	310	&	301	&$	1.49 	_{	-0.08 	}^{+	0.08 	}$&	0.30 	&	281/219	\\
--	&	311	-	332	&	320	&$	1.94 	_{	-0.13 	}^{+	0.18 	}$&	--	&	--	\\
--	&	335	-	391	&	355	&$	1.83 	_{	-0.15 	}^{+	0.16 	}$&	--	&	--	\\
130615A	&	356	-	368	&	362	&$	0.62 	_{	-0.09 	}^{+	0.09 	}$&	0.00 	&	764/659	\\
--	&	368	-	383	&	375	&$	0.64 	_{	-0.09 	}^{+	0.09 	}$&	--	&	--	\\
--	&	384	-	400	&	391	&$	0.75 	_{	-0.09 	}^{+	0.09 	}$&	--	&	--	\\
--	&	401	-	421	&	411	&$	0.85 	_{	-0.10 	}^{+	0.09 	}$&	--	&	--	\\
--	&	422	-	451	&	436	&$	1.00 	_{	-0.09 	}^{+	0.09 	}$&	--	&	--	\\
--	&	453	-	501	&	474	&$	1.15 	_{	-0.09 	}^{+	0.09 	}$&	--	&	--	\\
--	&	505	-	591	&	540	&$	1.37 	_{	-0.12 	}^{+	0.12 	}$&	--	&	--	\\
130925A1	&	997	-	1007	&	1002	&$	0.46 	_{	-0.10 	}^{+	0.10 	}$&	1.59 	&	 874/862	\\
--	&	1008	-	1019	&	1013	&$	0.51 	_{	-0.10 	}^{+	0.10 	}$&	--	&	--	\\
--	&	1020	-	1031	&	1025	&$	0.71 	_{	-0.11 	}^{+	0.11 	}$&	--	&	--	\\
--	&	1032	-	1043	&	1037	&$	0.59 	_{	-0.12 	}^{+	0.12 	}$&	--	&	--	\\
--	&	1044	-	1057	&	1050	&$	0.77 	_{	-0.12 	}^{+	0.12 	}$&	--	&	--	\\
--	&	1058	-	1086	&	1071	&$	0.86 	_{	-0.10 	}^{+	0.10 	}$&	--	&	--	\\
--	&	1087	-	1131	&	1107	&$	0.85 	_{	-0.10 	}^{+	0.10 	}$&	--	&	--	\\
130925A2	&	5020	-	5031	&	5026	&$	0.51 	_{	-0.10 	}^{+	0.10 	}$&	1.59 	&	 1189/1218	\\
--	&	5032	-	5042	&	5037	&$	0.58 	_{	-0.10 	}^{+	0.10 	}$&	--	&	--	\\
--	&	5043	-	5060	&	5051	&$	0.66 	_{	-0.09 	}^{+	0.09 	}$&	--	&	--	\\
--	&	5061	-	5083	&	5071	&$	0.80 	_{	-0.09 	}^{+	0.09 	}$&	--	&	--	\\
--	&	5084	-	5111	&	5097	&$	1.05 	_{	-0.09 	}^{+	0.09 	}$&	--	&	--	\\
--	&	5112	-	5162	&	5135	&$	1.24 	_{	-0.09 	}^{+	0.09 	}$&	--	&	--	\\
--	&	5163	-	5249	&	5203	&$	1.33 	_{	-0.10 	}^{+	0.10 	}$&	--	&	--	\\
131030A	&	116	-	125	&	120	&$	0.90 	_{	-0.09 	}^{+	0.09 	}$&	0.61 	&	907/769	\\
--	&	126	-	138	&	132	&$	1.23 	_{	-0.10 	}^{+	0.10 	}$&	--	&	--	\\
--	&	140	-	160	&	149	&$	1.51 	_{	-0.09 	}^{+	0.09 	}$&	--	&	--	\\
--	&	160	-	181	&	170	&$	1.75 	_{	-0.08 	}^{+	0.09 	}$&	--	&	--	\\
--	&	182	-	207	&	193	&$	2.02 	_{	-0.09 	}^{+	0.09 	}$&	--	&	--	\\
--	&	208	-	218	&	213	&$	1.99 	_{	-0.12 	}^{+	0.12 	}$&	--	&	--	\\
--	&	219	-	225	&	222	&$	2.09 	_{	-0.16 	}^{+	0.16 	}$&	--	&	--	\\
140430A	&	222	-	233	&	227	&$	1.76 	_{	-0.15 	}^{+	0.16 	}$&	0.91 	&	319/298	\\
--	&	233	-	239	&	236	&$	2.06 	_{	-0.18 	}^{+	0.19 	}$&	--	&	--	\\
--	&	239	-	246	&	242	&$	2.49 	_{	-0.19 	}^{+	0.20 	}$&	--	&	--	\\
--	&	247	-	257	&	251	&$	2.78 	_{	-0.19 	}^{+	0.20 	}$&	--	&	--	\\
--	&	259	-	274	&	266	&$	3.22 	_{	-0.27 	}^{+	0.28 	}$&	--	&	--	\\
140506A	&	359	-	372	&	365	&$	2.64 	_{	-0.15 	}^{+	0.15 	}$&	0.64 	&	772/567	\\
--	&	373	-	384	&	378	&$	2.38 	_{	-0.11 	}^{+	0.11 	}$&	--	&	--	\\
--	&	385	-	394	&	389	&$	2.57 	_{	-0.14 	}^{+	0.14 	}$&	--	&	--	\\
--	&	395	-	417	&	405	&$	2.76 	_{	-0.12 	}^{+	0.12 	}$&	--	&	--	\\
--	&	417	-	440	&	428	&$	3.11 	_{	-0.15 	}^{+	0.15 	}$&	--	&	--	\\
--	&	441	-	483	&	459	&$	3.40 	_{	-0.16 	}^{+	0.16 	}$&	--	&	--	\\
140512A	&	144	-	150	&	147	&$	0.16 	_{	-0.11 	}^{+	0.11 	}$&	0.20 	&	346/304	\\
--	&	151	-	156	&	153	&$	0.51 	_{	-0.14 	}^{+	0.14 	}$&	--	&	--	\\
--	&	157	-	163	&	160	&$	0.64 	_{	-0.15 	}^{+	0.15 	}$&	--	&	--	\\
--	&	164	-	173	&	169	&$	0.88 	_{	-0.17 	}^{+	0.17 	}$&	--	&	--	\\
--	&	175	-	187	&	180	&$	1.00 	_{	-0.19 	}^{+	0.19 	}$&	--	&	--	\\
140709A	&	189	-	192	&	190	&$	1.97 	_{	-0.32 	}^{+	0.33 	}$&	0.28 	&	187/187	\\
--	&	193	-	199	&	196	&$	1.94 	_{	-0.28 	}^{+	0.29 	}$&	--	&	--	\\
--	&	200	-	206	&	203	&$	2.05 	_{	-0.24 	}^{+	0.25 	}$&	--	&	--	\\
--	&	207	-	216	&	211	&$	2.32 	_{	-0.29 	}^{+	0.31 	}$&	--	&	--	\\
--	&	218	-	235	&	225	&$	2.80 	_{	-0.36 	}^{+	0.40 	}$&	--	&	--	\\
141031A	&	1104	-	1109	&	1106	&$	1.09 	_{	-0.28 	}^{+	0.29 	}$&	0.12 	&	 263/220	\\
--	&	1111	-	1139	&	1124	&$	1.10 	_{	-0.13 	}^{+	0.13 	}$&	--	&	--	\\
--	&	1141	-	1164	&	1152	&$	1.05 	_{	-0.19 	}^{+	0.19 	}$&	--	&	--	\\
--	&	1167	-	1187	&	1177	&$	0.95 	_{	-0.20 	}^{+	0.20 	}$&	--	&	--	\\
--	&	1191	-	1225	&	1206	&$	1.06 	_{	-0.20 	}^{+	0.20 	}$&	--	&	--	\\
--	&	1229	-	1256	&	1241	&$	0.95 	_{	-0.29 	}^{+	0.29 	}$&	--	&	--	\\
--	&	1262	-	1578	&	1264	&$	0.51 	_{	-0.96 	}^{+	0.84 	}$&	--	&	--	\\
141130A	&	329	-	336	&	332	&$	2.22 	_{	-0.36 	}^{+	0.39 	}$&	0.20 	&	110/126	\\
--	&	338	-	352	&	344	&$	2.64 	_{	-0.40 	}^{+	0.59 	}$&	--	&	--	\\
--	&	355	-	376	&	364	&$	2.77 	_{	-0.35 	}^{+	0.37 	}$&	--	&	--	\\
--	&	380	-	404	&	391	&$	3.26 	_{	-0.43 	}^{+	0.46 	}$&	--	&	--	\\
--	&	409	-	445	&	425	&$	3.11 	_{	-0.41 	}^{+	0.48 	}$&	--	&	--	\\

\enddata
\centering
\end{deluxetable}

\clearpage
\begin{deluxetable}{l|ccccc}
\footnotesize
\tablewidth{0pc}
\tablecaption{Curvature effect model fits to the evolution behaviors of flux and spectral index in the steep decay phase of the X-ray flares in our sample.}
\tablehead{
\colhead{GRB} & \colhead{$F_{E,0}$}& \colhead {$\hat{\beta}$}& \colhead {$E_{c,0}$}& \colhead {$t_{c}$}&\colhead{$\chi^{2}/dof$} \\
\colhead{} & \colhead{($10^{-9}\rm{erg}\cdot\rm{cm^{-2}}\cdot\rm{s^{-1}}\cdot \rm keV^{-1}$)}&\colhead {}& \colhead {(\rm keV)} & \colhead{(\rm{s})} & \colhead{}
}
\startdata
050502B	&	1.77 	$\pm$	0.05 	&	1.69 	$\pm$	0.04 	&	5.72 	$\pm$	0.24 	&	171.71 	$\pm$	3.38 	&	1.32 	\\
060111A	&	2.28 	$\pm$	0.07 	&	1.51 	$\pm$	0.04 	&	7.76 	$\pm$	0.24 	&	121.13 	$\pm$	2.17 	&	3.03 	\\
60124	&	2.74 	$\pm$	0.16 	&	1.04 	$\pm$	0.06 	&	2.90 	$\pm$	0.12 	&	123.25 	$\pm$	4.58 	&	1.80 	\\
060904B	&	16.46 	$\pm$	0.72 	&	0.93 	$\pm$	0.18 	&	2.54 	$\pm$	0.24 	&	65.39 	$\pm$	1.41 	&	3.98 	\\
60929	&	1.37 	$\pm$	0.07 	&	1.10 	$\pm$	0.06 	&	4.62 	$\pm$	0.16 	&	121.01 	$\pm$	2.00 	&	8.31 	\\
070520B	&	2.33 	$\pm$	0.12 	&	1.56 	$\pm$	0.06 	&	3.28 	$\pm$	0.10 	&	138.59 	$\pm$	2.71 	&	3.76 	\\
70704	&	3.58 	$\pm$	0.19 	&	1.17 	$\pm$	0.06 	&	3.96 	$\pm$	0.12 	&	129.90 	$\pm$	2.39 	&	4.42 	\\
80928	&	1.03 	$\pm$	0.08 	&	0.68 	$\pm$	0.10 	&	3.42 	$\pm$	0.26 	&	67.73 	$\pm$	3.13 	&	1.36 	\\
091130B	&	9.05 	$\pm$	0.52 	&	0.28 	$\pm$	0.53 	&	1.64 	$\pm$	0.46 	&	86.13 	$\pm$	2.34 	&	2.09 	\\
100619A	&	2.73 	$\pm$	0.10 	&	1.12 	$\pm$	0.05 	&	3.22 	$\pm$	0.06 	&	253.92 	$\pm$	3.21 	&	6.51 	\\
100704A	&	24.07 	$\pm$	0.99 	&	1.11 	$\pm$	0.16 	&	2.74 	$\pm$	0.20 	&	49.50 	$\pm$	1.13 	&	3.32 	\\
100802A	&	2.14 	$\pm$	0.07 	&	1.23 	$\pm$	0.05 	&	4.62 	$\pm$	0.16 	&	242.57 	$\pm$	4.42 	&	2.93 	\\
100902A	&	52.50 	$\pm$	1.65 	&	1.12 	$\pm$	0.15 	&	2.88 	$\pm$	0.24 	&	43.86 	$\pm$	0.76 	&	6.64 	\\
110801A	&	2.40 	$\pm$	0.41 	&	1.46 	$\pm$	0.08 	&	2.84 	$\pm$	0.14 	&	113.27 	$\pm$	5.56 	&	1.24 	\\
110820A	&	2.59 	$\pm$	0.56 	&	0.31 	$\pm$	0.28 	&	4.32 	$\pm$	1.08 	&	23.98 	$\pm$	1.19 	&	6.01 	\\
121027A	&	1.06 	$\pm$	0.14 	&	1.81 	$\pm$	0.05 	&	5.38 	$\pm$	0.14 	&	264.63 	$\pm$	8.92 	&	1.64 	\\
121211A	&	5.96 	$\pm$	0.23 	&	0.10 	$\pm$	0.08 	&	1.22 	$\pm$	0.02 	&	179.31 	$\pm$	6.44 	&	2.22 	\\
121229A	&	1.26 	$\pm$	0.06 	&	1.35 	$\pm$	0.07 	&	3.74 	$\pm$	0.16 	&	130.88 	$\pm$	3.18 	&	3.69 	\\
130131A	&	2.02 	$\pm$	0.16 	&	2.00 	$\pm$	0.05 	&	5.92 	$\pm$	0.20 	&	57.87 	$\pm$	1.75 	&	3.24 	\\
130615A	&	1.05 	$\pm$	0.04 	&	0.88 	$\pm$	0.06 	&	3.94 	$\pm$	0.16 	&	156.82 	$\pm$	3.20 	&	2.34 	\\
130925A1	&	5.18 	$\pm$	0.14 	&	0.66 	$\pm$	0.04 	&	3.56 	$\pm$	0.04 	&	157.63 	$\pm$	1.35 	&	13.79 	\\
130925A2	&	6.61 	$\pm$	0.15 	&	0.39 	$\pm$	0.04 	&	2.38 	$\pm$	0.02 	&	247.51 	$\pm$	1.45 	&	18.70 	\\
131030A	&	47.79 	$\pm$	1.86 	&	1.46 	$\pm$	0.14 	&	4.54 	$\pm$	0.54 	&	54.61 	$\pm$	1.40 	&	1.76 	\\
140430A	&	7.25 	$\pm$	0.64 	&	2.25 	$\pm$	0.07 	&	3.42 	$\pm$	0.12 	&	51.52 	$\pm$	1.36 	&	2.48 	\\
140506A	&	11.43 	$\pm$	1.38 	&	1.96 	$\pm$	0.97 	&	2.04 	$\pm$	1.14 	&	129.17 	$\pm$	21.95 	&	1.21 	\\
140512A	&	3.11 	$\pm$	0.26 	&	0.10 	$\pm$	0.54 	&	2.54 	$\pm$	0.12 	&	43.60 	$\pm$	2.83 	&	0.88 	\\
140709A	&	10.43 	$\pm$	1.76 	&	1.82 	$\pm$	0.11 	&	3.00 	$\pm$	0.12 	&	33.55 	$\pm$	1.16 	&	1.90 	\\
141031A	&	0.73 	$\pm$	0.05 	&	1.44 	$\pm$	0.07 	&	5.86 	$\pm$	0.16 	&	180.30 	$\pm$	6.71 	&	1.39 	\\
141130A	&	0.66 	$\pm$	0.35 	&	2.11 	$\pm$	0.17 	&	2.04 	$\pm$	0.14 	&	167.56 	$\pm$	10.17 	&	1.78 	\\
\hline
\enddata
\centering
\end{deluxetable}
\clearpage
\begin{sidewaystable}
\scriptsize
\centering
\caption{$R_{\rm X}$ and $\Gamma_{\rm X}$ for flares with known redshift}
\begin{tabular}{ccccccccccc}
\hline
\hline
GRB  & z & $\alpha$ & $\beta$ & $E_{p}$ &model& $\chi^{2}/dof$ &$F_{p}$&  $L_{\rm X, p}$&$ \Gamma_{\rm X}$ &$R_{\rm X}$  \\
 & & & & $({\rm keV})$& & & $(10^{-9}\rm{erg}\cdot\rm{cm^{-2}}\cdot\rm{s^{-1}})$& $(10^{48}\rm{erg}\cdot\rm{s^{-1}})$& &$(10^{15}\rm{cm})$ \\
\hline
060124	&	2.297	&$	2	_{	-0.08	}^{+	0.08	}$&	--					&	--			&	power-law	&	1.01	&$	6.78	\pm	0.16	$&$	286	\pm	6.74	$&$	55.60 	\pm	3.43 	$&$		6.93 	\pm	0.89 		$	\\
060904B	&	0.703	&$	0.93 	_{	-0.09 	}^{+	0.09 	}$&	--					&	--			&	band	&	1.08	&$	50.24	\pm	1.1	$&$	113	\pm	2.47	$&$	40.54 	\pm	2.89 	$&$		3.79 	\pm	0.55 		$	\\
080928	&	1.692	&$	0.39 	_{	-0.14 	}^{+	0.13 	}$&	--					&	--			&	power-law	&	1.01	&$	4.17	\pm	0.13	$&$	82.6	\pm	2.58	$&$	36.46 	\pm	2.73 	$&$		2.01 	\pm	0.32 		$	\\
110801A	&	1.858	&$	2.61 	_{	-0.10 	}^{+	0.11 	}$&	--					&	--			&	power-law	&	0.99	&$	5.07	\pm	0.13	$&$	127	\pm	3.25	$&$	42.15 	\pm	2.96 	$&$		4.23 	\pm	0.63 		$	\\
121027A	&	1.773	&$	1.19 	_{	-0.13 	}^{+	0.14 	}$&	--					&	--			&	power-law	&	1.05	&$	2.72	\pm	0.06	$&$	60.5	\pm	1.33	$&$	32.79 	\pm	2.56 	$&$ 6.16 	\pm	0.98 	 $	\\
121211A	&	1.023	&$	-0.91 	_{	-0.51 	}^{+	0.38 	}$&$	-2.49 	_{	-0.27 	}^{+	0.11 	}$&$	2.02	\pm	1.71	$&	band	&	1.98	&$	14.57	\pm	0.22	$&$	82.8	\pm	1.25	$&$	36.49 	\pm	2.71 	$&$	 7.08 	\pm	1.08  	$	\\
121229A	&	2.707	&$	1.12 	_{	-0.17 	}^{+	0.17 	}$&	--						& --			&	power-law	&	1.14	&$	3.19	\pm	0.06	$&$	200	\pm	3.76	$&$	49.24 	\pm	3.21 	$&$	 5.14 	\pm	0.68 $	\\
130925A1	&	0.347	&$	0.46 	_{	-0.10 	}^{+	0.10 	}$&	--					&	--			&	power-law	&	1.01	&$	22	\pm	0.11	$&$	9.1	\pm	0.05	$&$	17.19 	\pm	1.71 	$&$ 2.08 	\pm	0.41  $	\\
130925A2	&	0.347	&$	0.51 	_{	-0.10 	}^{+	0.10 	}$&	--					&	--			&	power-law	&	0.98	&$	23	\pm	0.1	$&$	9.5	\pm	0.04	$&$	17.46 	\pm	1.73 	$&$	 3.36 	\pm	0.67 $	\\
131030A	&	1.293	&$	-0.91 	_{	-0.34 	}^{+	0.46 	}$&$	-3.19 	_{	-0.39 	}^{+	0.24 	}$&$	2.13	\pm	2.04	$&	band	&	1.15	&$	106.67	\pm	3.81	$&$	1080	\pm	38.8	$&$	87.51 	\pm	4.45 	$&$ 10.94 	\pm	1.15 $	\\
140430A	&	1.6	&$	2.75 	_{	-0.14 	}^{+	0.14 	}$&	--					&	--			&	power-law	&	0.96	&$	15.71	\pm	0.35	$&$	271	\pm	6.04	$&$	54.61 	\pm	3.40 	$&$ 3.55 	\pm	0.45  $	\\
140506A	&	0.889	&$	2.64 	_{	-0.15 	}^{+	0.15 	}$&	--					&	--			&	power-law	&	1.36	&$	16.1	\pm	0.48	$&$	64.6	\pm	1.93	$&$	33.53 	\pm	2.60 	$&$ 4.61 	\pm	1.06  $	\\
140512A	&	0.725	&$	-0.42 	_{	-0.50 	}^{+	0.73 	}$&$	-1.61 	_{	-0.07 	}^{+	0.05 	}$&$	5.14	\pm	4.51	$&	band	&	1.45	&$	70.14	\pm	4.64	$&$	170	\pm	11.3	$&$	46.60 	\pm	3.27 	$&$ 3.29 	\pm	0.51  $	\\
\hline
\end{tabular}
\end{sidewaystable}
\end{document}